\documentclass[a4paper]{article}
\pdfoutput=1
\usepackage{jheppub}
\usepackage{url}
\usepackage{graphicx}
\usepackage{subcaption}
\usepackage{booktabs}
\usepackage{amsmath}
\usepackage{amssymb}
\usepackage{amstext}
\usepackage{amsfonts}
\usepackage{amsthm}
\usepackage{hyperref}
\usepackage{appendix}
\usepackage{physics}
\usepackage[utf8]{inputenc}
\usepackage{soul}
\usepackage{hhline}
\usepackage{comment}
\usepackage{color}
\usepackage{bm}
\usepackage[normalem]{ulem}
\usepackage{subcaption}
\newcommand\eea{\end{eqnarray}}
\newcommand\bea{\begin{eqnarray}}
\def\vec{\boldsymbol}
\def\l{\left(}
\def\r{\right)}
\def\OO{\mathcal{O}}
\newcommand{\eV}{{\, {\rm eV}}}
\def\Lt{t_{\rm trans}}
\def\mA{m_{A'}}

\title{\large{Searching for String Bosenovas with Gravitational Wave Detectors}}
\date{\today}
\author[a]{Dawid Brzeminski,}
\author[a]{Anson Hook,}
\author[b]{Junwu Huang,}
\author[a]{and Clayton Ristow}

\affiliation[a]{Maryland Center for Fundamental Physics, University of Maryland, College Park, MD 20742, USA}
\affiliation[b]{Perimeter Institute for Theoretical Physics, 31 Caroline St.~N., Waterloo, Ontario N2L 2Y5, Canada}

\emailAdd{dbrzemin@umd.edu}
\emailAdd{hook@umd.edu}
\emailAdd{jhuang@perimeterinstitute.ca}
\emailAdd{cristow@umd.edu}

\abstract{We study the phenomenology of a string bosenova explosion in vector superradiance clouds around spinning black holes, focusing on the observable consequences in gravitational wave detectors and accelerometers. During the superradiance growth of a dark photon cloud, the dark electromagnetic field might reach a critical field strength, when a network of dark photon strings is produced via a superheated phase transition. 
These dark photon strings will then absorb the energy in the background gauge fields and get ejected from the cloud, with total energy of the string network as large as the total rotational energy of the spinning black hole. 
In this paper, we study the subsequent evolution of this dense string network, and the resulting observational consequences depending on the unknown string tension, or almost equivalently, the ratio between the quartic and the gauge coupling in the Abelian Higgs model. Strings with large tension will dissipate into gravitational waves, detectable over a wide range of frequencies, from $\sim {\rm nHz}$ near supermassive blackholes, to $\gtrsim 10 {\rm MHz}$ around stellar mass black holes. This is the first known source of high frequency gravitational waves, unconstrained by cosmological observations. The strain of this gravitational wave can be larger than $10^{-14}$ at low frequencies, lasting for longer than typical duration of experiments. Small tension strings, whose string networks can have total lengths as large as $10^{40}\,{\rm km}$, can travel to the earth with appreciable rate from any black hole in the Milky Way and interact with earth based accelerometers. If the Standard Model particles are directly charged under the dark photon, e.g.~$U(1)_{\rm B-L}$, this interaction leads to an acceleration of Standard Model particles that is \emph{independent} of the coupling strength. We work out the spectral density of this acceleration, and project that modern accelerometers and equivalence principle tests can be sensitive to the passing of these strings.}

\begin{document}

\maketitle

\section{Introduction}

A dark photon, a light gauge boson from a Higgsed dark $U(1)$ symmetry, is one of the simplest extensions to the Standard Model. The dark photon is motivated as a low energy consequence of string theory~\cite{Goodsell:2009xc,Cicoli:2011yh}.  It has been suggested as a dark matter candidate~\cite{Nelson:2011sf,Arias:2012az,Hui:2016ltb}, and can serve as a mediator of interactions to a dark sector~\cite{Pospelov:2008jk,Hall:2009bx,Essig:2011nj,Knapen:2017xzo}. The mass range of interest for the dark photon spans many orders of magnitude, contributing to a diverse experimental program searching for light vectors~\cite{Chaudhuri:2014dla,Baryakhtar:2018doz,SuperCDMS:2019jxx,FUNKExperiment:2020ofv,SENSEI:2020dpa,Chiles:2021gxk,Cervantes:2022yzp,Brzeminski:2022rkf,Alonso-Alvarez:2023tii}. 

Recently, it was found that at large dark photon (field and energy) densities there is a phase transition which leads to the production of a huge number of vortices~\cite{East:2022ppo,East:2022rsi}. In the case of early universe dark photon production, such a phase transition often leads to almost all of the energy density in dark photons dissipating into gravitational wave~\cite{East:2022rsi}, and the depletion of dark photon dark matter (see also~\cite{Cyncynates:2023zwj}).
On the other hand, in the case of late time dark photon production, similar dynamics still occur but instead lead to spectacular signatures.  In the case of dark photon superradiance around spinning black holes~\cite{Baryakhtar:2017ngi}, the phase transition leads to periodic string bosenova events: the periodic production of vortex strings around spinning black holes~\cite{East:2022ppo,East:2022rsi}.   
These strings can emit gravitational waves with a large enough amplitude to be detectable as well as interact directly with detectors in the lab~\cite{East:2022rsi}. In this paper, we work out some of the signatures outlined in~\cite{East:2022rsi} in more detail. 

The dynamics of string production can be described with the well-known Abelian Higgs model in the gravitational background of a Kerr black hole with mass $M$ and spin $a_*\sim 1$
\begin{equation}\label{eq:abelianhiggs}
    \mathcal{L} = \left|D_{\mu} \Phi\right|^2 -\frac{1}{4}F'^{\mu\nu}F'_{\mu\nu}+\frac{\lambda}{4}\left(|\Phi|^2-v^2\right)^2,
\end{equation}
where $A'_{\mu}$ is the dark photon and $\Phi$ is complex scalar. $D_{\mu} \equiv \partial_{\mu} - i g A'_{\mu}$ is the covariant derivative, $v$ is the vev of the Higgs field, $\lambda$ is the quartic coupling and $g$ is the gauge coupling of the dark $U(1)$. The dark photon mass is $m_{A'} = \sqrt{2} g v$ while the Higgs field has a mass of $\lambda^{1/2} v$.  The system dynamics depend on 3 independent dimension-less combinations of these parameters~\footnote{We have adopted natural units $\hbar = c =1$}: 
\begin{itemize}
    \item $\alpha \equiv G M m_{A'}$, which determines the growth of the superradiance cloud
    \item $\lambda/g^2$, which determines the number of strings produced per phase transition event
    \item $G\mu \approx \pi G v^2$, which determines the evolution of the string network\footnote{The string tension $\mu$ is $\pi  v^2$ multiplied by the log of the string core radius over the smaller of the string separation or the dark photon Compton wavelength, which is a time dependent function. For the rest of the discussion, we neglect this weak time dependence.} 
\end{itemize}
The scenario we consider works as follows. For any spinning black hole, if there is a dark photon with mass such that $\alpha = G M m_{A'} \lesssim 1$, a superradiant instability is triggered, and a superradiant cloud of dark photon grows exponentially in amplitude and energy density, extracting energy and angular momentum from the Kerr black hole~\cite{Baryakhtar:2017ngi}. In the limit of $ \lambda/g^2 \rightarrow\infty$, the cloud grows until the black hole's angular velocity drops below the dark photon mass, saturating the superradiance condition~\cite{Baryakhtar:2017ngi}. The cloud dissipates slowly through gravitational wave (and potentially also electromagnetic radiation~\cite{Baryakhtar:2017ngi,Siemonsen:2022ivj}). 

On the other hand, for finite $ \lambda/g^2$ (even as large as $\sim 10^{40}$), the superradiant growth of the cloud ends when a critical energy density of $B_{\rm sh}^2 \simeq \lambda v^4 $ is reached in the cloud.  At this point a superheated phase transition takes place and converts almost all of the energy in the classically rotating superradiance cloud into a network of dark photon strings. These strings, once produced, start to interact with each other and get ejected by the remaining dark electromagnetic field as well as by the repulsive interactions between nearby strings. The ejected strings will continue to interact with each other and the background dark photon radiation, releasing gravitational waves as the system expands at close to the speed of light. The gravitational radiation power, and also the fraction of energy that goes into gravitational wave depends mainly on the string tension in Planck units $G\mu$. 

The expanding network of strings will go through three stages, depending on both $\lambda/g^2$ and $G\mu$. The first is the tightly coupled regime, where the interaction between the emitted dark photon radiation and strings  as well as the string-string interaction keeps the string-dark photon network in a tightly coupled equilibrium. 
After this network of strings has expanded for a significant amount of time, the network enters a weakly interacting regime where the string-dark photon radiation interactions are no longer strong enough to keep the string network in equilibrium with the dark photon radiation.  In this limit the string network decays, releasing gravitational wave and dark photon radiation, while the remaining strings continuing to interacting with each other.  Eventually, the string density is so small that even the string-string interaction is no longer efficient and the network disappears through gravitational wave and dark photon emission. During the whole period, gravitational waves of a variety of frequencies will be emitted, leaving unique observable consequences.  Meanwhile, if their lifetime is long enough, the strings themselves may arrive at earth giving observable direct detection signals.

A particularly useful combination of these parameters is the fraction of black hole spin energy $ \mathcal{F}_s\sim (G \mu) (\lambda/g^2)/\alpha^5 a_*$ released in each string bosenova event. The inverse $\mathcal{F}_s^{-1}$ is the number of bosenova cycles a black hole will go through.  For a fixed $\mathcal{F}_s$, the two parameters $\lambda/g^2$ and $G\mu$ are inversely proportional to each other. Whereas the parameter $\alpha$ has to reside in a narrow range of $10^{-3}$ to $  1$ for superradiance to occur, $\lambda/g^2$ and $G\mu$ can both take values that differ by many orders of magnitude. Two of the most motivated possibilities are $\lambda/g^2 \gtrsim 1$, which occurs when the gauge coupling and the quartic coupling are both very small and of similar origin, or $\lambda/g^2 \gg 1$, which occurs if the quartic coupling are much larger than the small dark gauge coupling.  In the former case, a small number of high tension strings are produced, which leads to a burst of gravitational waves following the string formation. In the latter case, a large number of low tension strings are produced and that expand like a ``fireball'' of strings for a long time, leading to strings that can interact directly with earth based detectors. In both cases, there are observable effects in various gravitational wave detectors and accelerometers. 

Gravitational wave signals from the network of strings can show up in a wide range of frequencies, from as low as $10^{-8}$ Hz from supermassive black holes, to as high as $10$ MHz from stellar mass black holes, with peak amplitude $h_c \sim 10^{-19}$ in the aLIGO band, and $h_c \sim 10^{-15}$ in the LISA/PTA frequency range, lasting for a time of order milliseconds to thousands of years. These periodic gravitational wave bursts are within the sensitivity of current and future gravitational wave detectors, proposed to cover a wide range of frequencies. 
Of particular note is that due to string-string interactions, a string bosenova event leads to the formation of a large abundance of both long and short strings with lengths that are parametrically longer and smaller than the size of the black hole (by a factor of $(\lambda/g^2)^{1/3}$ and $(\lambda/g^2)^{-1/2}$, respectively).  In particular, the short strings emits gravitational waves at frequencies much larger than $1/G M$ of the black hole giving a compelling production mechanism of high frequency gravitational waves at amplitudes large enough to be experimentally relevant.  This gives the {\it only} simple scenario where very high frequency gravitational waves can come from known stellar objects.  The prospect of a detectable signal at high frequency gravitational wave detectors is as a result, of particular interest~\cite{Arvanitaki:2012cn, Domcke:2022rgu,Berlin:2023grv}. 

For direct detection of strings, we consider the particularly intriguing case where Standard Model particles are directly charged under the dark $U(1)$, as in the case of a light $B-L$ gauge boson.  Similar to interactions between charged particles and magnetic monopoles, the interaction between a $B-L$ gauge boson string and Standard Model (SM) particles has a strength that is independent of the gauge coupling strength $g$, as the dark photon string carries a unit magnetic flux of $2\pi/g$ while SM particles interact with a strength $g$. These strings leave a unique signature in gravitational wave detectors as well as equivalence principle violating tests, depending only on a single parameter $m_{A'}$, the mass of the gauge boson. This is the {\it second} target for lab search for topological defects~\cite{Pospelov:2012mt}, apart from the famous magnetic monopole~\cite{Cabrera:1982gz}.

The paper is organized as follows. In section~\ref{Sec: String Dynamics}, we discuss the production and evolution of the dark photon string network after a string bosenova. In section~\ref{Sec:GW} and~\ref{Sec:DirectImpact}, we work out the observational consequences of the string network from both the gravitational waves emitted as the network evolves and from direct detection of dark photon strings at earth-based detectors.

\section{The String Bosenova Cycle}
\label{Sec: String Dynamics}
In this section, we describe the string Bosenova cycle. We start with the formation of strings from the superradiance of the dark photons around a black hole and then describe the evolution of those strings after production until their inevitable disappearance due to decays into dark photons. 

\begin{figure}[h!]
\centering
\includegraphics[width=.5\linewidth]{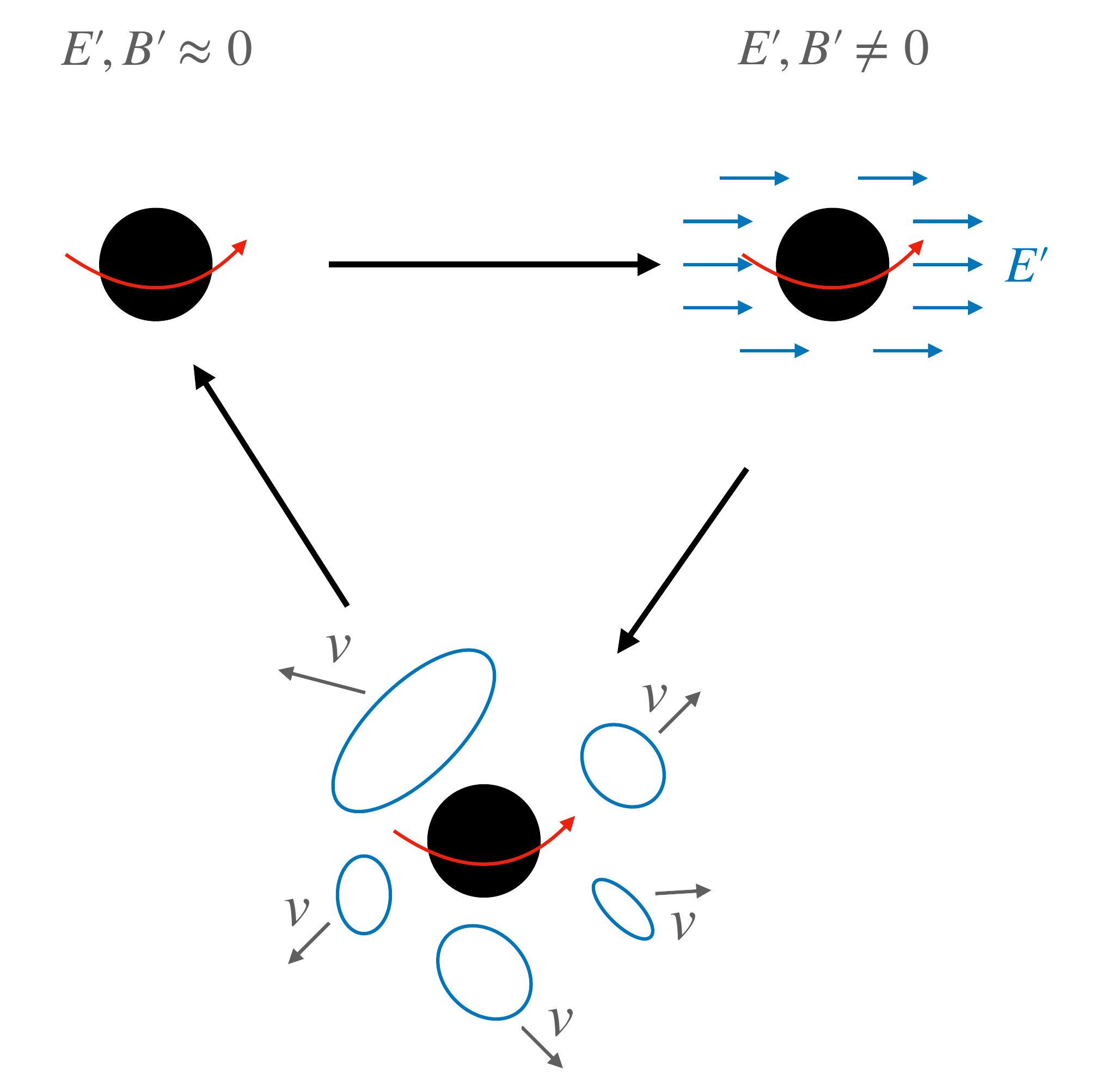}
\caption{A schematic representation of the life cycle of a spinning black hole.  Initially we begin with a spinning black hole.  Subsequently superradiance begins to form a dark photon cloud, that is, a coherent oscillating dark electric and magnetic field.  Finally, these dark fields become so large that it undergoes a phase transition producing many strings that discharge the dark fields and are ejected from the black hole.  After this event, the black hole is back to its initial state and the whole sequence begins anew.} 
\label{Fig: production}
\end{figure}

\subsection{Superradiant string production}
If the dark photon has a Compton wavelength that is comparable or slightly larger than the Schwarzschild radius of a spinning black hole, it will be produced through the superradient instability, where the dark photon amplitude grows exponentially, extracting energy and angular momentum from the black hole. The superradiant production of dark photons was studied in Ref.~\cite{Baryakhtar:2017ngi}, where it was found that the fastest growing mode was the orbital angular momentum $\ell=0$. It was found that the angular momentum in the dark photon cloud grows exponentially with rate 
\bea
\label{Eq: Superradiance Rate}
\Gamma_{\rm SR} \simeq 4 a_*(t) \alpha^6 m_{A'}
\eea
where $a_*$ is the spin of the black hole.  Naively, the cloud grows until all available angular momentum has been extracted from the black hole. The superradiant cloud subsequently slowly decay through gravitational wave (and possibly electromagnetic~\cite{Siemonsen:2022ivj}) emission, lasting for years or even centuries~\footnote{The higher $\ell$ modes of the superradiant cloud can also grow after the $\ell = 0$ depletes~\cite{Arvanitaki:2014wva,Baryakhtar:2017ngi}. We neglect these contributions since they are subdominant.}.

However, as was demonstrated in Ref.~\cite{East:2022rsi}, if the field strength ever reaches a critical value of $B'_{\rm sh}=v^2 \sqrt{\lambda}$ before the black hole's angular momentum depletes, the superradiant cloud undergoes a super-heated phase transition into a network of dark photon strings. 
This network of strings quickly exits the vicinity of the black hole, allowing for a new superradiant cloud to form and the cycle to repeat~\footnote{While not found in the numerical simulation of Ref.~\cite{East:2022rsi}, it is possible that there is a transitory period where a string becomes attached to the black hole and inhibits superradiance.  The string would spin out string length until it detaches from the black hole and superradiance can begin again~\cite{Deng:2023cwh}.}.  Using the fact that the superradiant cloud has a radius $\sim (m_{A'}\alpha)^{-1}$, one can compute the fraction of the black holes spin extracted in this process, $\mathcal{F}_s$, and the number of bursts, $N_b$, to be 
\bea
\label{Eq: Spin Fraction}
\mathcal{F}_s \approx \frac{\lambda}{g^2}\frac{G \mu}{\alpha^5 a_*} \qquad N_b = \mathcal{F}_s^{-1} .
\eea
From the exponential growth rate in Eq.~\ref{Eq: Superradiance Rate}, we can compute the time, $t_{\rm SH}$, it takes for the superraniance cloud to reach $B'_{\rm sh}$
\bea
\label{Eq: Superheating Time}
t_{\rm SH}=\frac{1}{4 a_*^0\alpha^6 m_{A'}}\ln\l a_*^0GM_{\rm BH}^2\frac{\mathcal{F}_s}{1-\mathcal{F}_s}\r
\eea
where $a_*^0$ is the initial spin of the black hole.  We can also define the superradiant life time $\tau_{\rm BH}\equiv t_{\rm SH}/\mathcal{F}_s$ as the total time the black hole undergoes superradiance. This will be important to assess how many black holes in our galaxy are undergoing superradiance and thus are capable of producing strings. 

Note that integrating out the heavy radial mode of $\Phi$ also induces a quartic $F^4$ interaction of dark photon, which, similar to the light scalar case~\cite{Baryakhtar:2020gao,Gruzinov:2016hcq}, leads to mixing between different superradiance levels and emission of semi-relativistic dark photon through $3$ to $1$ processes. In the limit of large $\lambda/g^2$, these modifications become important only at dark photon field strengths parametrically larger than the critical field for string production.  As a result, these mixing effects will be unimportant for our purposes.

\subsection{String Evolution}

\begin{figure}[h!]
\centering
\includegraphics[width=.7\linewidth]{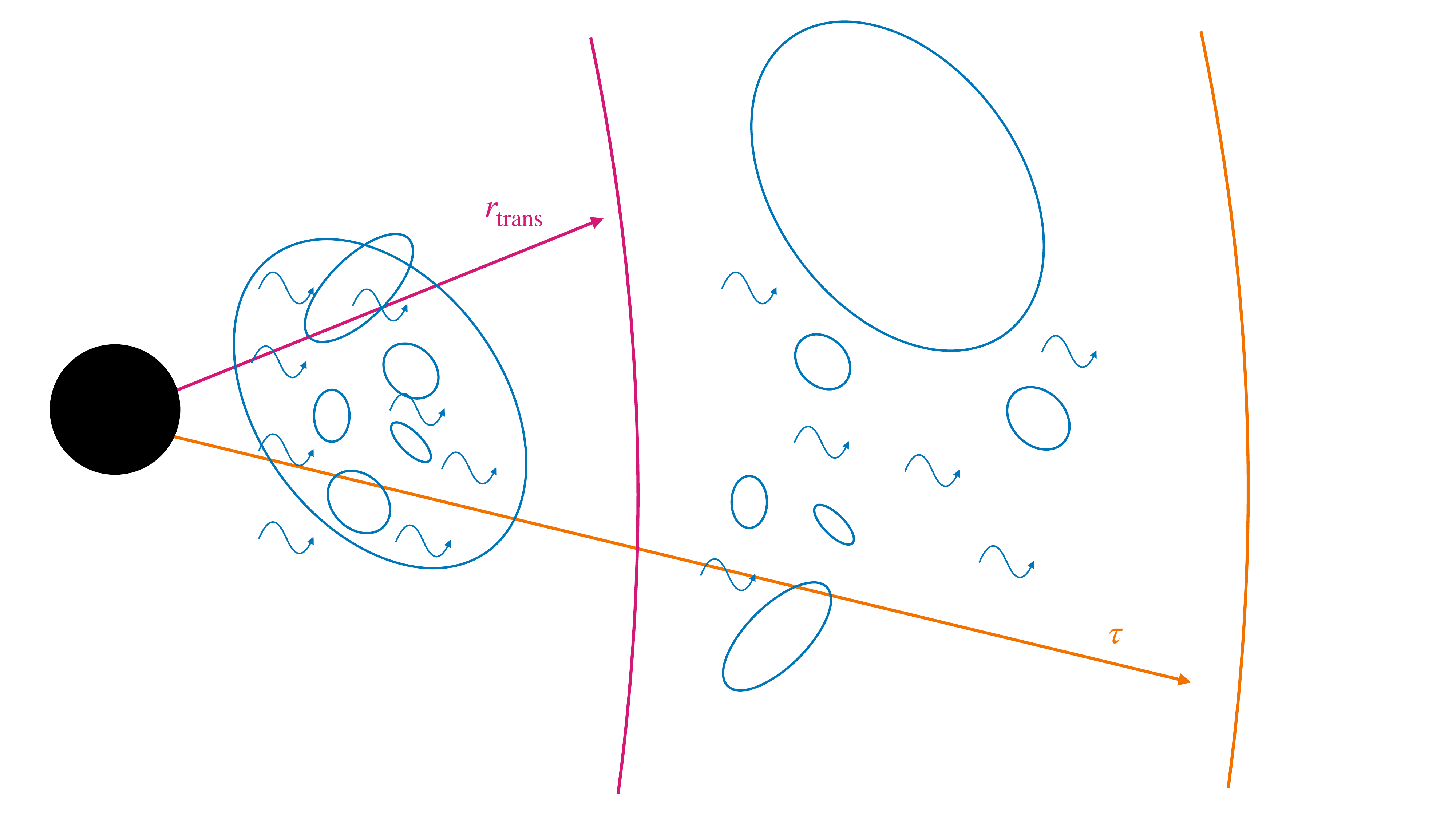}
\caption{A schematic representation of how the produced strings propagate away from the black hole.  Initially after production and ejection, the dark photons and dark strings are tightly coupled with each other.  Constant string formation from string-string interaction and emission and absorption of dark photon radiation due to string-radiation interaction lead to a dynamical equilibrium during the initial expansion.  Eventually at a radius $r_{\rm trans}$, the tightly coupled approximation fails and the strings and dark photons decouple from each other.  In this regime, the dark photons radiation freely propagate, while the strings might remain interacting with each other. These strings radiate dark photons and gravitational waves and eventually disappear at a radius $\tau$, the string network lifetime.} 
\label{Fig: propagation}
\end{figure}

Now let us discuss the dynamics of the expanding string network following the string bosenova burst. 
In principle, the string dynamics can be obtained by numerically solving the equations of motion for the Higgs field and dark photon.  Simulating these field theory equations are complicated, but necessary for a full understanding of the tightly coupled regime, that is when backreaction of the dark photon radiation bath on the strings is important.  In the decoupled regime, where the strings and the dark photon radiation decouple from each other, the string equations of motion can instead be used to solve for the string dynamics.  In practice, both of these numerical simulations might be too complicated and resource demanding and we will instead need to make educated guesses on what their dynamics may be.

As field theory simulations are well established~\cite{East:2022rsi}, we will not elaborate on these simulations but instead point out another possibility, by considering the strings as described by the Nabu-Goto action, with couplings to a dark photon. In this language, the string dynamics can be described by strings coupled to a massive 1-form gauge fields, with the following action~\cite{Polchinski:1998rq}: 
\bea
   \mathcal{S} = -\mu \int d^2 x \sqrt{ \det g_{ab}} - \int d^4 x - \frac{1}{4m_{A'}^2} H_{\alpha \mu \nu}H^{\alpha \mu \nu} + \epsilon_{ab} G_{\mu\nu} \partial^a X^{\mu}\partial^b X^{\nu} ,
\eea 
where $g^{ab}=\partial^a X_{\mu}\partial^b X_{\nu} g^{\mu\nu}$. At linear order, this results in the following equations of motions which describes string-gauge field interactions:
\begin{align}
    \mu \Box_s X^{\mu} &= H_{ \mu \alpha\nu} J^{\alpha \nu}\nonumber\\
    \frac{1}{m_{A'}^2} \partial^{\alpha} H_{\alpha\mu\nu} -G_{\mu\nu}&= \frac{1}{g} J_{\mu\nu} 
\end{align}
where $G_{\mu\nu}$ is an 1-form gauge field, $ H_{\alpha\mu\nu}  = \partial_{\alpha} G_{\mu\nu} - \partial_{\mu} G_{\alpha\nu}+\partial_{\nu} G_{\alpha\mu} $ is its field strength, and $J_{\mu\nu} = \epsilon_{a b} \partial^{a} X_{\mu}\partial^{b} X_{\nu}$ is the corresponding current. $X^{\mu} (\sigma, \tau)$ describes the motion of the string as a function of the world sheet coordinates. The Roman letters denote world sheet coordinates while the Greek letters the 3+1D spacetime coordinates. $\Box_s$ is the d'Alembert operator in world sheet coordinate. The full non-linear action also describes the formations of small loops from long strings apart from the interactions between the background gauge fields with the strings. In particular, the important dynamics of emission of dark photon radiation (transverse and longitudinal modes) from the (short) strings, the formation of small string loops from long wiggly string loops, and the backreaction of the emitted radiation on the long string loops play a key role in the string networks evolution once it is formed. 

While dedicated numerical simulations are needed to fully determine the evolution of this expanding network, it seems clear that the system undergoes two stages of evolution: a tightly coupled period and a weakly/non-interacting period.  Let us first give a broad strokes overview of what we suspect these periods look like before getting into a more quantitative discussion.

In the decoupled regime, the emitted dark photons and the strings are decoupled from each other~\footnote{Note however that the strings will still intersect with themselves and each other.}.  The strings dilute away, intersecting with each other and evaporating due to the emission of dark photons and gravitational waves, while the emitted dark photons and gravitational waves propagate freely.
Meanwhile, in the tightly coupled regime, the dark photon amplitude is so large that interactions between the strings and dark photons are unavoidable, e.g. the dark photons can accelerate strings as occurs around the black hole that produced them. This region necessitates a more qualitative estimate based on numerical simulations of similar situations.

A rough sketch of our string evolution as a function of time is shown in Fig.~\ref{Fig: propagation}.  Initially, the string network is produced in the tightly coupled regime.  The fireball of strings and radiation proceeds to expand at the speed of light.  Eventually, the densities involved fall until the strings and radiation fall out of equilibrium and enter the decoupled regime. The string-string interaction might proceed to decouple during the subsequent evolution. Finally, these strings either reach the Earth or evaporate before reaching the Earth.

In analogy to cosmic simulations of local cosmic strings, the string length is most likely dominated by the longest strings after a long period of evolution after a string bosenova event. While we keep most of the discussions in the following subsections general, we will also present estimates of the signal strength in the long string (IR) dominated regime to motivate the explicit computations in the next sections.  
\subsubsection{Tightly coupled regime}

After the black hole produces the strings via a superheated phase transition, the string length can lie anywhere in the range $m_{A'}^{-1}(\lambda/g_D^2)^{-1/2} < l <(\alpha m_{A'})^{-1}$, where the upper and lower bounds come from the total size of the production region and the average distance between strings respectively~\footnote{Depending on how one estimates the minimum string length, either via total string length or via the strength of the overlapping B fields, you can get either a $-1/2$ or $-1/4$ respectively for the minimum string length.}.  If the strings and gauge bosons did not interact, these strings would then radiate into dark photon longitudinal modes.  Smaller string loops ($l \lesssim (m_{A'})^{-1}$) would radiate with power $P_{\rm DP}\propto \mu$. Longer strings ($l > (m_{A'})^{-1}$) radiate at rates that are suppressed by $(m_{A'} l)$ with the exact scaling still under debate. Namely $P_{\rm DP}\propto  \mu (m_{A'} l)^{-1}$, where $\mu\sim \pi v^2$ is the string tension and the $-1$ power in the exponential can vary as it depends on the exact mechanism of energy loss~\cite{Vachaspati:1984gt,Allen:1991bk, Roshan:2024qnv}.  These strings would, as a result, emit away all their energy in dark photon longitudinal modes on time scales of $l$ and  possibly $m_{A'} l^2$ respectively and disappear. 

However, at least initially, this picture is not correct, as it does not account for the backreaction of the emitted radiation on the strings. In fact, if all the strings became radiation in such short amount of time, the background field strengths would rise back to close to the superheating field strength $B'_{\rm sh}$, far above the critical field $B'_{\rm c1} = g v^2$ where strings are energetically favored over background magnetic fields.  At fields above $B'_{\rm c1}$, strings would absorb energy from the background radiation instead of emitting radiation.  Such behavior is observed in the 3D dark photon dark matter simulation in Ref.~\cite{East:2022rsi}, where it was observed that the short strings live for lifetimes much longer than their erstwhile lifetime.

It is clear then that, initially, the expanding network of strings is tightly coupled, and all emitted radiation get reabsorbed by the string network, very much like a strongly interacting plasma. The network of strings and its surrounding radiation is described by a string length distribution function $F_s(l, t)$ similar to that of the scaling solution found for cosmological strings.  Meanwhile the radiation energy density is described by a distribution $F_r(k, t)$. The distribution functions are normalized such that 
\bea \label{Eq: string length normalization}
    \int \mathrm{d} l \, F_s(l, t) \, l = L_{\rm tot} (t)
\eea
and 
\bea~\label{eq:radiation}
    \int \mathrm{d} \log k \, \rho (k,t) \, F_r(k, t) \simeq B_{\rm c1}^{\prime 2}
\eea
where $\rho(k,t)$ is the energy density of dark photons.
Here, we make the reasonable assumption that the ambient energy in dark photons is of order ${B^{\prime 2}_{\rm c1}}$, which is the cusp between strings absorbing or emitting radiation.  We expect the ambient radiation density to remain near this critical point as the string network expands while the string network itself decreases in density as it expands. Note that Eq.~\ref{eq:radiation} is not required for the rest of the discussions in the paper, our conclusions hold as long as the energy density in the dark photon radiation does not exceed the energy density of the string network in the tightly coupled regime.

In analogy with the early universe case, we expect the string network to evolve towards a distribution like cosmic strings in the scaling regime.  Cosmic strings in the scaling regime have distributions that are functions of $l/t$ with factors of $1/t$ making of the rest of the dimensions.  We therefore assume that the length distribution of strings takes a fixed shape that scales with the radius $r(t)$ of the expanding network 
\bea \label{eq: length distribution}
   F_s(l, t)= f(l/r(t))\frac{L_{\rm tot} (t)}{r^2(t)}.
\eea
The factor of $L_{\rm tot} (t)$ is a normalization constant while the $1/r^2(t)$ is there for dimensional reasons.  Finally $f(x)$ is normalized such that $\int dx \,x \, f(x)=1$. Without a detailed numerical simulation of this string network, the form of $f(x)$ is a priori unknown. For simplicity, we take a power law $f(x)\sim x^q$ . For $q>-2$, the total length will be concentrated in strings of length on the order of the size of the network ($l\sim r(t)$) while for $q<-2$ , the total length will be concentrated in strings of smallest length. It will be of particular interest to determine this scaling dimension $q$ as a function of time with dedicated simulation.

The total length $L_{\rm tot}(t)$ can decrease as the network expands due to emission of gravitational waves as well as due to the emission of dark photons needed to maintain $B' \sim B'_{\rm c1}$ as the volume increases.  The emitted gravitational wave power of a single string is $P_{\rm GW}=\Gamma_{\rm GW}G\mu^2$, where $\Gamma_{\rm GW} \sim 50$~\cite{Vilenkin:1981bx,Blanco-Pillado:2011egf,Blanco-Pillado:2013qja,Blanco-Pillado:2017oxo}.  One can solve for the total length after a time $t$ due to emission of gravitational waves, 
\bea
\label{Eq: Total Length Change Strong Regime}
\frac{L_{\rm tot}(t)}{L_{\rm tot}(0)}=(\alpha m_{A'}r(t))^{-\beta \Gamma_{\rm GW} G\mu} .
\eea
where $\beta=\int dx f(x)$ is a parameter that characterizes the length distribution of strings, that can be matched on to dedicated numerical simulations. We expect $\beta$ to be only weakly time dependent.  If the string network is long string dominated, $\beta$ is an $\OO(1)$ constant.  We also take the strings to move at the speed of light, that is, $r(t) = t$ in deriving this equation. Here, the total string length $L_{\rm tot}(0) = \frac{\lambda}{g^2}/\alpha^3 m_{A'} $ right after the string bosenova.   
In the parameter region of interest $G\mu \ll 1$ and the emission of gravitational waves is a highly subdominant source of energy exchange between the strings and radiation, and only a small fraction of the total energy of the tightly coupled system at any time.  Despite this, these emitted gravitational waves can be observable at present day gravitational wave detectors.

The dominant energy loss of the string network will come from the emission of dark photons needed to maintain $B' \sim B'_{\rm c1}$ as the volume increases.
Eventually, due to energy conservation, it is impossible to maintain this large of a magnetic field and the tightly coupled regime ends. 
Around the same time, the magnetic fields of different strings also stop overlapping with each other. The volume occupied by the expanding network of strings, at this point, is
\bea
\label{Eq: Volume Transition}
    V_{\rm transition} \sim r^3(t_{\rm trans})= \frac{\lambda}{g^2}\left(\frac{1}{\alpha m_{A'}}\right)^3 \qquad   r(t_{\rm trans})=\left(\frac{1}{\alpha m_{A'}}\right)\l \frac{\lambda}{g^2}\r^{1/3}.
\eea
As a result of the tightly coupled nature of the string, the maximum string length can be of order $(\lambda/g^2)^{1/3}$ larger than their initial length of the string at production.  The minimum string length presumably also evolves as a function of time. The minimum string length can be found to be the average distance between strings, in this case $1/m_{A'}$, consistent with our previous statement that the strings stop overlapping.

\subsubsection{Decoupled Regime} \label{subsec:decoupled}
The string network now enters the decoupled regime, where the emitted radiation in both gravitational wave and dark photon can both escape the system. Strings emit dark photon radiation and gravitational wave radiation with rates 
\bea
\label{Eq: Total Power}
    P = \Gamma_{\rm GW} G\mu^2 +  \Gamma_{\rm DP}\mu \min\l1,(m_{A'}l)^{-1}\r .
\eea
The $\min\l1,(m_{A'}l)^{-1}\r$ enforces that the dark photon power is different for strings of length shorter than $m_{A'}^{-1}$ than for strings of length longer than $m_{A'}^{-1}$. As before, the factor of $(m_{A'}l)^{-1}$ depends on the energy loss mechanism but suffices for the parameteric estimates that we do in this section.  While gravitational wave emission is the same for strings of all lengths, radiation into dark photons is slowest for strings of the longest possible length. For our network exiting the tightly coupled regime, these strings have length $l \leq r(t_{\rm trans})$~\footnote{Percolation theory involving strings often ends up with infinite strings, e.g. cosmologically produced strings end up having $\sim 80\%$ of their string length in an infinite string.  Therefore it is possible that $\mathcal{O}(1)$ of the total string length in our scenario is in a single string that is wrapped up very tightly in a ball.  Given the large uncertainties in our estimates, we expect that such a possibility would not change our results by more than an $\mathcal{O}(1)$ number, but may result in interesting experimental signatures separate from what we consider.}.  Therefore, for all strings, 
\bea
\frac{P_{\rm GW}}{P_{\rm DP}} = \frac{\Gamma_{\rm GW}}{\Gamma_{\rm DP}} G\mu m_{A'} l \lesssim G\mu m_{A'}(\alpha m_{A'})^{-1}\l\frac{\lambda}{g^2}\r^{1/3} = \l\frac{\lambda}{g^2}\r^{-2/3}\alpha^4 a_* \mathcal{F}_s\ll 1,
\eea
suggesting that the lifetime of strings is completely determined by the emission into dark photon radiation. One can solve Eq.~\ref{Eq: Total Power} to find the life time $\tau(l_0)$ of some string with initial length $l_0$
\bea
\label{Eq: String Lifetime}
\tau(l_0) \sim \begin{cases}
  \frac{m_{A'}^2l_0^2+1}{2m_{A'}\Gamma_{\rm DP}} & \text{ if}\quad m_{A'} l_0 > 1 \\
  \frac{l_0}{\Gamma_{\rm DP}} & \text{ if} \quad m_{A'}l_0<1
\end{cases}
\eea

The behavior of strings in the decoupled regime is very different depending on if the string distribution is UV or IR dominated.  If the string distribution is UV dominated, the strings will dominantly have a small size of $1/\mA$ and will quickly decay in time $1/\mA$.  Thus UV dominated string distributions have an extremely short lived decoupled regime.

IR dominated string distributions have $\tau > t_{\rm trans}$ and are thus much more interesting.  Let us first check that the strings continue to self interact.  A string of length $l$ will intersect the typical string (of length $r(t)$) within its lifetime as long as $n \sigma \tau > 1$.  The number density of typical strings is $n \sim L_{\rm tot}/r(t)^4$ while the cross section scales as $\sigma \sim l r(t)$.  As the lifetime of the string is long, $\tau \sim \mA l^2$, we find that all strings with length
\bea
L_\text{dec} > r(t) (\mA L_{\rm tot})^{-1/3}
\eea
will continue to intersect.  Meanwhile, the distance between strings is $L_{\rm d} \sim r(t) (r(t)/L_{\rm tot})^{1/2}$.  The minimal string length is likely one of these two, though because the string distribution is IR dominated, this minimal length does not affect the evolution or the observational consequence.

The evolution of the total length of the string network can be obtained from emission of dark photon:
\bea
\frac{d L_{\rm tot}}{dt} \sim \frac{1}{\mA l} \frac{L_{\rm tot}}{l},
\eea
where $l \sim r(t)$ is the average string length.  The solution to this equation is 
\bea\label{Eq:IRdecoupled}
L_{\rm tot}(t) = L_{\rm tot}(0) e^{-\frac{1}{\mA} (1/t_{\rm trans} - 1/t)},
\eea
 suggesting that emission of dark photons does not significantly change the total string length during string expansion.  

Because the total length does not change appreciably, the string network expands until $r(t) \sim L_{\rm tot}$, when the entirety of the string network is in $\mathcal{O}(1)$ strings.  At this point, the string oscillates and very slowly decays with a lifetime of order $\tau \sim \mA L_{\rm tot}^2$. Depending on the ratio $\lambda/g^2$, the time $L_{\rm tot} \propto \lambda/g^2$ and $\mA L_{\rm tot}^2\propto (\lambda/g^2)^2 $ can both be longer than the age of the Universe, in which case the evolution has not completed.

\subsubsection{Prospects for Detection}

For a fixed $\mathcal{F}_s$, there are two important regimes under which to consider this evolution: large $\lambda/g^2$ and extremely small $G\mu$, and moderate  $G\mu$ and $\lambda/g^2$. Note that, when taking these limits, we must always have $\lambda/g^2 > 1$ in order for strings to form and $G\mu<1$ so that the symmetry breaking scale of the radial mode is smaller than the Planck scale. This discussion is intended to offer an estimate of the signal rate and size, and more detailed computations are presented in sections~\ref{Sec:GW} and~\ref{Sec:DirectImpact}.

\paragraph{Long lived strings: Large $\lambda/g^2$}
In this regime we have $\lambda/g^2\gg 1$ and $G\mu\ll 1$. Due to the small $G\mu$, the gravitational radiation can be easily neglected through out the evolution. The network of strings conserve total string length during the strongly interacting regime. The string network exits this regime after
\bea
\label{Eq: ttrans}
    t_{\rm trans}  \sim\frac{1}{\alpha m_{A'}} \left(\frac{\lambda}{g^2}\right)^{1/3} 
\eea
 Interestingly, if this time scale is longer than the time between string bosenova explosion, $t_{\rm sh}$, the strings from the different string bosenova events will overlap and the string network will stay in the tightly coupled regime for a factor of $\sqrt{t_{\rm trans}/t_{\rm sh}}$ longer.

Due to the large $\lambda/g^2$ the strings entering the decoupled regime will have comparatively large lengths due to the long expansion period. As a consequence, the lifetime of these strings can be quite long, especially if the string network is IR dominated. For the longest strings of initial length $l\sim (\alpha m_{A'})^{-1}(\lambda/g^2)^{1/3}$, they have a lifetime 

\bea
    \tau \sim m_{A'} l^2 \sim \left(\frac{\lambda}{g^2}\right)^{2/3} \frac{1}{\alpha^2 m_{A'}} =2 \, \times \, 10^5 \,{\rm Gyr} \left(\frac{\lambda/g^2}{10^{30}}\right)^{2/3}\left(\frac{10^{-13} \,{\rm eV}}{m_{A'}}\right) \left(\frac{10^{-2}}{\alpha}\right)^2,
\eea
potentially much longer than the age of the universe. These strings actually continue to grow in length after the string-DP interaction decouples, due to efficient string-string intersection, which makes this lifetime even longer. These strings can easily make their way to us here on Earth where they can be detected by ground-based detectors. The prospects for this direct detection are detailed in Section~\ref{Sec:DirectImpact}.

\paragraph{Short lived strings: Moderate $G \mu$ and $\lambda/g^2$} On the other hand, there is the regime of moderately small $G\mu$ and moderately large $\lambda/g^2$. Since gravitational radiation scales with $G\mu$, the strings in this regime can produce a significant amount of gravitational wave energy. We can estimate the energy deposited into gravitational waves, $E_{\rm GW}$ by summing the energy deposited in the tightly and weakly coupled regimes. In the tightly coupled regime, this energy can be found to be 
\begin{equation} \label{Eq: tightly coupled}
E_{\rm GW}^{\rm tight}
 \simeq \mu L_{\rm tot}(0)
 \frac{\beta\Gamma_{\rm GW}G\mu}{3}\ln\l\frac{\lambda}{g^2}\r.
\end{equation}
Taking $\beta \sim f_\text{typical} t_{\rm trans}$ in terms of the typical frequency $f_\text{typical} \sim 1/l_\text{typical}$ of the emitted gravitational waves, this equation matches a more precise derivation given in the next section.
For a fixed $\mathcal{F}_s\sim \OO(1)$, the GW energy for IR dominated length distributions is
\bea 
\frac{E_{\rm GW}^{\rm tight}}{M_{\rm BH}}\sim 10^{-12} \l \frac{a_* \mathcal{F}_s}{1} \r \l\frac{m_{A'}}{10^{-13}\text{ eV}}\frac{M_{\rm BH}}{10 M_\odot}\r\l\frac{\Gamma_{\rm GW} G\mu}{10^{-10}}\r\ln\l\frac{\lambda}{g^2} \r.
\eea

As emphasized in the previous subsection, the behavior in the decoupled regime depends heavily on if the distribution is UV or IR dominated.  For UV dominated distributions, the energy emitted in the decoupled regime can be estimated from the lifetime of the typical sized strings at the time of exiting the tightly coupled regime.
\bea
E_{\rm GW, UV}^{\rm decoupled} &=&\Gamma_{\rm GW}G\mu^2 L_{\rm tot}(0),
\eea
a factor of $\beta \sim \mA t_{\rm trans}$ smaller than that of the tightly coupled regime.  This suppression is due to the fact that instead of living a long time $\sim t_{\rm trans}$, they are able to decay away in a much shorter amount of time $\sim 1/\mA$.

For IR dominated length distributions, there are gravitational waves emitted from when the string distribution is still evolving and gravitational wave emission from the final string with $l \sim L_{\rm tot}(0)$.  When the string distribution is still evolving, as can be seen in Eq.~\ref{Eq:IRdecoupled}, the emission is dominated at the scale $t_{\rm trans}$.  The gravitational wave emission from this part of the evolution can be estimated to be
\begin{eqnarray}\label{eq:EGW}
E_{\rm GW}^{\rm decoupled} &=&\Gamma_{\rm GW}G\mu^2\tau(l_\text{typical}) \frac{L_{\rm tot}(0)}{l_\text{typical}} \\
&\sim& \Gamma_{\rm GW}G\mu^2m_{A'} L_{\rm tot}(0) t_{\rm trans} \lesssim \mu L_{\rm tot}(0)\Gamma_{\rm GW}G\mu\alpha^{-1} \l\frac{\lambda}{g^2}\r^{1/3} \nonumber,
\end{eqnarray}
where we assumed the IR dominated case, and the factor of $L_{\rm tot}(0)/l_\text{typical}$ is a rough estimate of the number of strings emitting power.  As before, for a fixed $\mathcal{F}_s\sim \OO(1)$, we have
\bea 
\frac{E_{\rm GW, IR}^{\rm ssc}}{M_{\rm BH}}\sim 10^{-10} \l \frac{a_* \mathcal{F}_s}{1} \r 
\l\frac{\Gamma_{\rm GW} G\mu}{10^{-10}}\r\l\frac{\lambda}{g^2} \r^{1/3}.
\eea

Finally, there is the GW emission coming from the string with $l \sim L_{\rm tot}(0)$ 
at the end of the decoupled regime.  The energy emitted is
\begin{eqnarray}
E_{\rm GW, IR}^{\rm decoupled} =\Gamma_{\rm GW}G\mu^2 \mA L_{\rm tot}(0)^2,
\end{eqnarray}
and 
\begin{eqnarray}
\frac{E_{\rm GW, IR}^{\rm decoupled}}{M_{\rm BH}} \simeq 10^{-6} \l \frac{a_* \mathcal{F}_s}{1} \r^2 \l\frac{m_{A'}}{10^{-13}\text{ eV}}\frac{M_{\rm BH}}{10 M_\odot}\r^3 \l \frac{\Gamma_{\rm GW}}{\Gamma_{\rm DP}}\r.
\end{eqnarray}

Whereas the estimates presented in this subsection are very crude, it does suggest that gravitational wave radiation from both stages can be a significant portion of the black hole's energy, and therefore could produce a noticeable strain at gravitational wave detectors. 
The frequency spectrum of this signal and its strain are computed in Section~\ref{Sec:GW}\footnote{The analytical computation of the gravitational wave strain presented in this paper should be considered as an addition to the gravitational wave and dark photon radiation emitted during the initial burst of string formation, which can only be computed from simulations~\cite{East:2022ppo}}.

\section{Gravitational wave signature}\label{Sec:GW}

In this section, we estimate the sensitivity of gravitational wave detectors to the gravitational waves emitted by the string network following a bosenova event.  
The GW signal of string bosenova events comes from a combination of a large number of strings, and is, as a results, similar to that of a stochastic background. This frequency dependent excess power resembles the GW emitted by a string network in the early universe. However,
as described in section~\ref{Sec: String Dynamics}, the property of the string network following a string bosenova event differ from an early universe string network (in the scaling regime) in two main ways, due to the strong string-dark photon radiation interaction, as well as the very frequent string-string interaction, both as a result of the enormous density of the strings at production.
In what follows, we describe how to calculate the frequency dependence and the amplitude of the gravitational waves emitted by the string network, accounting for these important differences, both during the decoupled and tightly coupled phases.

An isolated circular string loop emits power at a rate~\cite{Vachaspati:1984gt,Allen:1991bk, Roshan:2024qnv}
\bea \label{Eq: Power per string}
\frac{d P_{\rm GW}(f,l)}{df} = \frac{\Gamma_{\rm GW} G \mu^2}{\zeta (n)} \sum_{j=1}^{\infty} j^{-n} \delta \l f-\frac{2 j}{l} \r.
\eea
where $\zeta (n)$ is the Riemann zeta function and the value of $n$ depends on whether the dominant source of gravitational waves are cusps ($n=4/3$), kinks ($5/3$) or kink-kink collisions ($n=2$). While our strings are not particularly circular, we will use this expression with $n=1$ to obtain order of magnitude estimates of our signal.  The main point is that a loop of size $l$ tends to emit gravitational waves at a frequency $f \sim 1/l$.  As such, the frequency distribution of the gravitational waves emitted by a conglomeration of strings is determined by the length distribution of the string, $F_s(l,\tau)$.  The emitted gravitational waves have a frequency distribution
\bea
\label{Eq: GW master formula}
\frac{d \rho_{\rm GW}}{d \log f} &=& \frac{1}{4 \pi d^2 \Delta t} \int\limits_{t = 0, l = 0}^{t = \infty, l = \infty} d t dl F_s(l,t) f \frac{d P_{\rm GW}(f,l)}{df}, 
\eea
where $d$ is the distance to the bosenova event and $\Delta t$ is the duration of gravitational wave emission.  
We will present our result in terms of the characteristic strain $h_c(f)$, defined via \cite{Moore:2014lga,Maggiore:2007ulw} 
\bea \label{eq: rho and strain}
\frac{d \rho_{\rm GW}}{d \log f} = \frac{\pi}{16 \Delta t\, G}   h_c^2(f)   f  .
\eea
Combining equation~\ref{Eq: GW master formula} and~\ref{eq: rho and strain}, we get
\bea \label{eq: characteristic strain}
h_c(f) = \frac{2 G \mu \sqrt{\Gamma_{\rm GW}}}{\pi d \sqrt{\zeta (n)}}
\sqrt{ \sum_{j=1}^{\infty} \frac{j^{-n}}{2j} \int\limits_{t = 0}^{t = \infty} dt  F_s \l \frac{2j}{f},t\r  \l \frac{f}{2j}\r^{-2}} . 
\eea
This is valid for signals that are shorter than the duration of typical experiments. For signals that last longer than the duration of typical experiments $t_{\rm exp}$, there is an extra suppression of $\sqrt{t_{\rm exp}/\Delta t}$\footnote{Technically speaking, $h_c$ is a feature of the signal and not the experiment and is thus unchanged for experiments shorter than the duration of the signal.  What actually changes is the equation for the ${\rm SNR}$, Eq.~\ref{Eq: SnR GW}, however this change is equivalent to suppressing $h_c$ by $\sqrt{t_{\rm exp}/\Delta t}$.}.
In what follows, we make an educated guess on the form of $F_{s}(l,t)$ for the decoupled and tightly coupled phases and use Eq.~\ref{eq: characteristic strain} to find the characteristic strain of the signal. 

Like many astrophysical sources of gravitational waves, our signal has strong time dependence.  However, instead of going from low frequencies to high frequencies, our signal will sweep from high frequencies to low frequencies as the cloud of strings expands.  The signal quickly sweeps through the high frequency part of the signal and slows down towards the lower frequency piece. The string-string interaction eventually decouples and the frequency of gravitational wave signal increases again.  As a result of this behavior, when we eventually plot the characteristic strain of the signal, one should keep in mind that any given experiment will only see a subset of frequencies at any given point in time.

\subsection{Tightly coupled regime} \label{sec: GW tightly coupled}

\begin{figure*}[t] 
\centering
\begin{minipage}{.48\textwidth}
    \centering
    \includegraphics[width=0.995\linewidth]{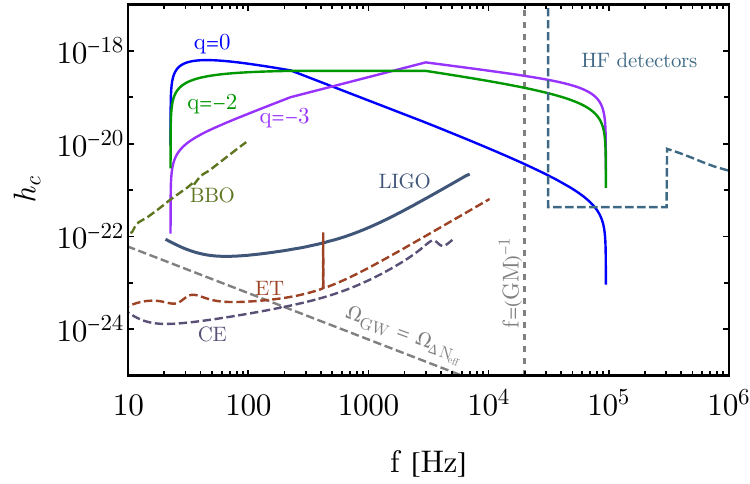}
\end{minipage}%
\hfill
\begin{minipage}{.48\textwidth}
    \centering
    \includegraphics[width=0.995\linewidth]{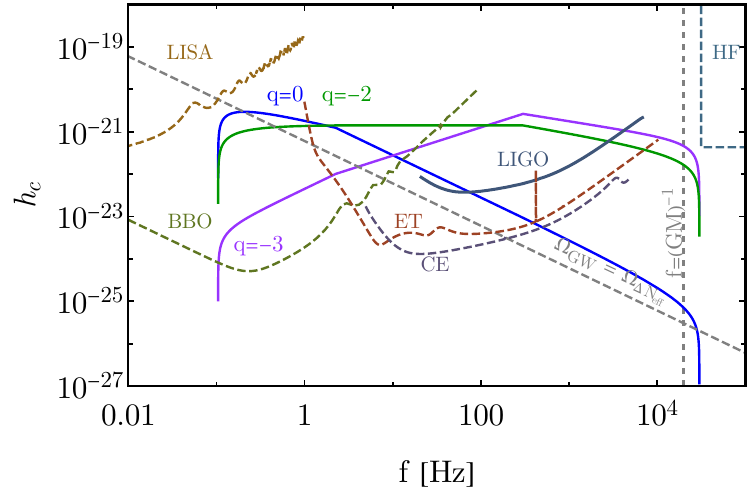}
\end{minipage}
\vskip\baselineskip

\begin{minipage}{.48\textwidth}
    \centering
    \includegraphics[width=0.995\linewidth]{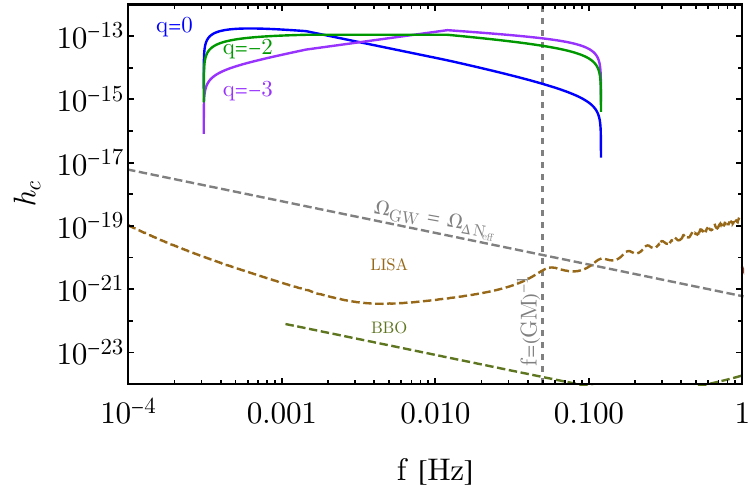}

\end{minipage}%
\hfill
\begin{minipage}{.48\textwidth}
    \centering
    \includegraphics[width=0.995\linewidth]{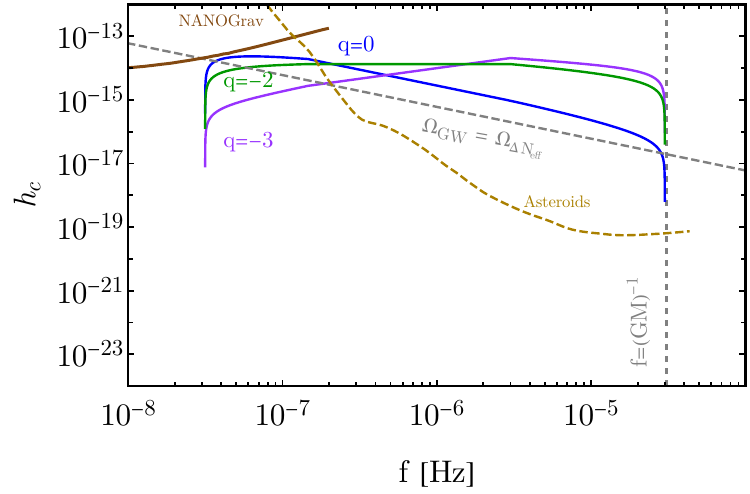}
\end{minipage}
\caption{An example of the strain spectrum generated during the tightly coupled regime as a function of the spectral index of the string network $q$, in comparison with the sensitivity of ongoing and proposed gravitational wave detectors across a wide range of frequencies: NANOgrav \cite{NANOGrav:2023ctt}, asteroids \cite{Fedderke:2021kuy},  LIGO-VIRGO-KAGRA \cite{KAGRA:2021tnv}, LISA \cite{Colpi:2024xhw}, BBO \cite{Cutler:2009qv}, ET \cite{Hild:2009ns}, CE \cite{Srivastava:2022slt}, levitating spheres \cite{Arvanitaki:2012cn}, axion haloscopes \cite{Domcke:2022rgu}. 
For string network that travels at speed $\lesssim c$, observationally, this signal can last for $\sim t_{\rm trans}$ before the strain starts to rise at low frequencies as parts of the string network start to enter the decoupled regime. Gravitational wave signals with large strain can be produced with frequencies larger than $1/GM$ of stellar mass black holes, producing a target for future high frequency gravitational wave detectors (HF detectors). The signal in this regime is a short burst with decreasing frequency, and dedicated searches shall be done with available data to look for this signal. The plots are made with the following set of parameters: a) top left, $M = 10 M_{\odot}$, $d = 1 \text{kpc}$, $\lambda/g^2 = 10^3$, $G \mu = 2.4 \times 10^{-9}$, $\mA = 10^{-12} \eV$;  b) top right, $M = 10 M_{\odot}$, $d = 1 \text{kpc}$, $\lambda/g^2 = 10^4$, $G \mu = 2.4 \times 10^{-15}$, $\mA = 10^{-13} \eV$; 
c) bottom left, $M = 4 \times 10^6 M_{\odot}$, $d = 8 \text{kpc}$, $\lambda/g^2 = 10^2$, $G \mu = 2.5 \times 10^{-7}$, $\mA = 4 \times 10^{-18} \eV$;  d) bottom right, $M = 6.5 \times 10^9 M_{\odot}$, $d = 16 \text{Mpc}$, $\lambda/g^2 = 10^2$, $G \mu = 2.7 \times 10^{-7}$, $\mA = 10^{-21} \eV$.}
    \label{fig: tightly coupled spectrum example}
\end{figure*}

We start by discussing the spectrum produced during the tightly coupled regime.
In this scenario, the strings are strongly interacting, and dark photons are being reabsorbed by strings in the cloud. Understanding how the length distribution of strings evolves in this regime would require dedicated simulation which is beyond the scope of this work. Instead, as explained in Sec. \ref{Sec: String Dynamics}, we work under the assumption that the length distribution of strings reaches a fixed shape, 
\bea \label{eq: length distribution SI}
   F_s(l, t)= f(l/r(t))\frac{L_{\rm tot} (0)}{r^2(t)} \l \Theta \l l- L_{\rm min}(t) \r -\Theta \l l- L_{\rm max}(t) \r \r,
\eea
where, $f(x) \sim x^q$. The index $q$ is generally a time dependent function. In particular, simulations in~\cite{East:2022rsi} suggest that the initial string network in the tightly coupled regime right after formation is UV dominated, while the string network towards the end of the expansion likely resembles a cosmic string network, where simulations generally find an IR dominated distribution~\cite{Gorghetto:2018myk,Gorghetto:2020qws,Buschmann:2019icd}. We account for this uncertainty by presenting the spectrum in the tightly coupled and decoupled regime for various values of $q$ that produce qualitatively different spectral shapes in figures~\ref{fig: tightly coupled spectrum example},~\ref{fig: decoupled spectrum example} and~\ref{fig: combined spectrum example} with a comparison shown in figure~\ref{fig: spectrum contributions example} in appendix~\ref{Appendix: exact results}.

We expect the fastest strings to travel at $v\approx c$, which results in the length of longest strings growing as $L_{\rm max}(t) = \frac{1}{\alpha \mA} + t$. Since the change of the total string length can be neglected as long as $G\mu \ll 1$, we take the length of shortest strings $L_{\rm min}(t)$ to be the average distance between the strings, $L_{\rm min}(t) = \frac{ L^{3/2}_{\text{max}}(t)}{\sqrt{\frac{\lambda}{g^2\alpha^2} L_{\rm max}(0) }}$ in this tightly coupled regime.

To simplify the analysis we further assume that each string emits GW primarily at the frequency $f=2/l$, corresponding to taking only the $j=1$ in equation~\ref{eq: characteristic strain}. 
After combining Eq. \eqref{eq: characteristic strain}, and Eq. \eqref{eq: length distribution SI} we get an expression for the characteristic strain 
\bea \label{eq: hc SI IR}
h_{c, {\rm tc}}(f) \approx \frac{ \sqrt{2 \Gamma_{\rm GW} L_{\rm tot} \Lt} G \mu}{ \pi d } \cdot
\begin{cases}  
 \l \frac{L_{\rm max}(0)}{ \Lt} \r^{\frac{1}{2}} \l \frac{f L_{\rm max}(0)}{2} \r^{\frac{-2-q}{2}}  \quad  & \frac{2}{L_{\rm min}(0)} > f > \frac{2}{L_{\rm max}(0)}
\\
\l \frac{f \Lt}{2} \r^{-\frac{1}{2}}  \quad & \frac{2}{L_{\rm max}(0)} > f > \frac{2}{\Lt} 
\\
0  \quad & f > \frac{2}{L_{\rm min}(0)} \text{ or } f < \frac{2}{\Lt} 
\end{cases} 
\eea
when $q>-1$, where the network dominated by long strings (IR dominated), and 

\bea \label{eq: hc SI UV}
h_{c, {\rm tc}}(f) \approx \frac{ \sqrt{2 \Gamma_{\rm GW} L_{\rm tot} \Lt} G \mu}{ \pi d }   \cdot \begin{cases}  
\l \frac{f}{2 \mA} \r^{-\frac{1}{3}}   & \frac{2}{L_{\rm min}(0)} > f > 2 \mA \\
   \l \frac{f}{2 \mA} \r^{\frac{-2-q}{2}}  & 2 \mA > f > \frac{2}{\Lt} 
\\
0 & f > \frac{2}{L_{\rm min}(0)} \text{ or } f < \frac{2}{\Lt} 
\end{cases}  
\eea
 when $q<-2$, where the network is dominated by short strings (UV dominated). Here, $ L_{\rm max}(t_{\rm trans})  = \frac{1}{\alpha m_{A'}} \left(\frac{\lambda}{g^2}\right)^{1/3} = \Lt$ is the size of the longest strings at the end of tightly coupled regime, and also roughly the duration of the tightly coupled regime.
The above expressions have been simplified for clarity and include leading contributions to the strain. Spectra shown in Fig. \ref{fig: tightly coupled spectrum example} are obtained from the full expressions that can be found in the Appendix \ref{Appendix: exact results}.

As can be seen in Fig. \ref{fig: tightly coupled spectrum example}, the high frequency tail of the distribution could potentially be seen by gravitational wave detectors sensitive to $\sim$ MHz frequencies.  This part of the spectrum is dominated by the emission of gravitational waves coming from when the strings are first produced around the black hole.  For UV dominated distributions, the maximum frequency $f_{\rm max} \sim 2\, m_{A'} \l \lambda/g^2 \r^{1/2}$ is related to the characteristic amplitude at that frequency by
\bea
h_c(f_{\rm max}) \approx 10^{-20} \l \frac{10 \, \text{kpc}}{d} \r \l \frac{a_\star}{1} \r \l \frac{\alpha}{0.1} \r^3 \l \frac{\mathcal{F}_s}{0.1} \r \l \frac{100 \, \text{kHz}}{f_{\rm max}} \r .
\eea
Depending on the details of the proposed gravitational wave detector, signals at frequencies as large as 10 MHz may be observable. The duration of the high frequency gravitational waves signal around this maximal frequency is short, lasting for a duration of $t_{\rm duration} \approx 1/\alpha m_{A'} = G M_{\rm BH}/\alpha^2$ ($f_{\rm max} t_{\rm duration} \sim (\lambda/g^2)^{1/2} / \alpha$), typically of order seconds or less. This maximal frequency of the signal decreases as a function of time, with
\begin{eqnarray}
    \frac{{\rm d} f}{{\rm d} t} = -\frac{3}{2}\frac{f^{5/3} \alpha m_{A'}}{f_{\rm max}^{2/3}}.
\end{eqnarray}

The qualitative features of the spectra in Fig. \ref{fig: tightly coupled spectrum example} are quite independent of the approximations made in deriving these results, and can be explained in the following way. During the tightly coupled regime, we expect strings to grow in size as they undergo interactions with themselves and the background field. The overall signal is determined by both the 
the number of strings emitting at a given frequency and the time they spend emitting at that frequency. The peak of the GW strain in Figures.~\ref{fig: tightly coupled spectrum example}, \ref{fig: decoupled spectrum example} and~\ref{fig: combined spectrum example} corresponds to the emissions from the typical strings in the network towards the end of the tightly coupled regime.  The typical strings have lengths of $l_{\rm typical} \sim t_{\rm trans}$ ($l_{\rm typical} \sim 1/m_{A'}$) if the strings are IR (UV) dominated, while the spectral index $q$ determines the powerlaw tail in the deep UV(IR). Detailed discussions about the qualitative behaviours away from the peak can be found in appendix~\ref{Appendix: exact results}.

Around the peak of the spectrum, the strain amplitude depends on the power emitted by a single string ($G\mu$), the number of strings contributing to the emission $L_{\rm tot}/l_{\rm typical}$, as well as the duration of the signal $\min (t_{\rm trans}, t_{\rm exp})$ . The different strings add up incoherently, leading to a strain $h_c$ that scales as the square root of $L_{\rm tot}/l_{\rm typical}$. The definition of $h_c$ is such that it also scales as the square root of $ \min (t_{\rm trans}, t_{\rm exp})\times l_{\rm typical}$, since the typical frequency is $2/l_{\rm typical}$. For the parameters presented in Figure~\ref{fig: tightly coupled spectrum example}, $t_{\rm trans}$ is always much smaller than the duration of typical GW experiments, and the peak strain is
\bea
h_{c,{\rm peak}} \sim 10^{-20} 
\l \frac{\lambda/g^2}{10^4} \r^{\frac{2}{3}} \l \frac{G\mu}{10^{-14}} \r \l \frac{M_{\rm BH}}{10 M_{\odot}} \r^{-2} \l \frac{\mA}{10^{-13} \eV} \r^{-3} \l \frac{d}{1 \text{kpc}} \r^{-1}. 
\eea
in agreement with the estimates in equation~\ref{Eq: tightly coupled}.

\subsection{GWs from the decoupled region} \label{sec: GW decoupled}
Now, let us move on to investigate the spectrum of GW produced after strings decoupled from dark photon radiation. Let us first briefly review the dynamics of the decoupled regime described in section \ref{subsec:decoupled}.  
Because the dynamics of UV  and IR dominated string distributions are different, we will discuss them separately.

\paragraph{IR dominated distributions : }

The string network is still coupled through string-string interaction. These interactions generally make the average string length continue to grow until string-string interactions decouple, while at the same time, the radiation from the string network reduces both the total and individual string length. For a string network that is IR dominated, the strings will continue to interact with each other and grow in length, similar to the tightly coupled regime. The rate of interaction for a string of length $l$ is:
\begin{equation}
    \Gamma_{\rm string} = \sigma_s n_s \simeq l L_{\rm tot}/V \simeq l L_{\rm tot}/ t^3,
\end{equation}
where $\sigma_s = l_1 l_2$ is the interaction cross section of two strings with length $l_1$ and $l_2$, and $n_s$ is the string number density. Comparing this to the string lifetime $m_{A'} l^2$, we get that the strings will at least interact once before they can dissipate into dark photon as long as $l > l_{\rm min} =  t \l m_{A'} L_{\rm tot} / \Gamma_{\rm DP} \r^{-1/3}$. Since typical strings in an IR dominated network has a length $l \sim t$, the string-string interaction is efficient and the string network evolves in the same way as in the tightly coupled regime, until there are only $\mathcal{O}(1)$ strings in the string network and the string has length $L_{\rm tot}$. During this evolution, the total length of the string
\begin{equation}
 \frac{{\rm d} L_{\rm tot}(t)}{{\rm d} t}  = \frac{\Gamma_{\rm DP} L_{\rm tot}(t)}{m_{A'} l^2}  \simeq \frac{\Gamma_{\rm DP} L_{\rm tot}(t)}{m_{A'} t^2},
\end{equation}
suggesting that the total string length is reduced by only a small fraction of $(m_{A'} t_{\rm trans})^{-1}$ in the IR dominated regime. For the remainder of the section, we neglect the reduction of the total string length during this period of evolution where the DP string interaction is decoupled while the string-string interaction is still efficient. This period of evolution ends at 
\begin{equation}
    t_{\rm ssc} \sim L_{\rm tot} \approx \frac{\lambda}{g^2} \frac{1}{\alpha^3 m_{A'}} = 7\times 10^4 \,{\rm s}\l \frac{\lambda/g^2}{10^5}\r \l \frac{0.1}{\alpha}\r^3 \l \frac{10^{-12} \,{\rm eV}}{m_{A'}}\r.
\end{equation}  
The strings now decays through dark photon radiation, with a lifetime of 
\begin{equation}
    \tau \simeq \frac{m_{A'} L_{\rm tot}^2}{2 \Gamma_{\rm DP}} \approx \l\frac{\lambda}{g^2} \r^2 \frac{1}{2 \Gamma_{\rm DP}\alpha^6 m_{A'}} = 20 \,{\rm yr} \l \frac{\lambda/g^2}{10^4}\r^2 \l \frac{0.1}{\alpha}\r^6 \l \frac{10^{-12} \,{\rm eV}}{m_{A'}}\r \l \frac{50}{\Gamma_{\rm DP}}\r,
\end{equation} 
at which point the signal would last longer than the duration of typical experiments.

\begin{figure*}[t] 
\centering
\begin{minipage}{.48\textwidth}
    \centering
    \includegraphics[width=0.995\linewidth]{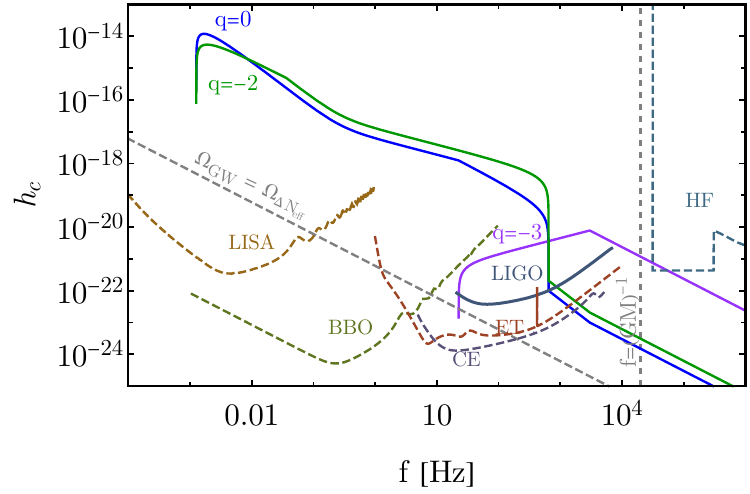}
\end{minipage}%
\hfill
\begin{minipage}{.48\textwidth}
    \centering
    \includegraphics[width=0.995\linewidth]{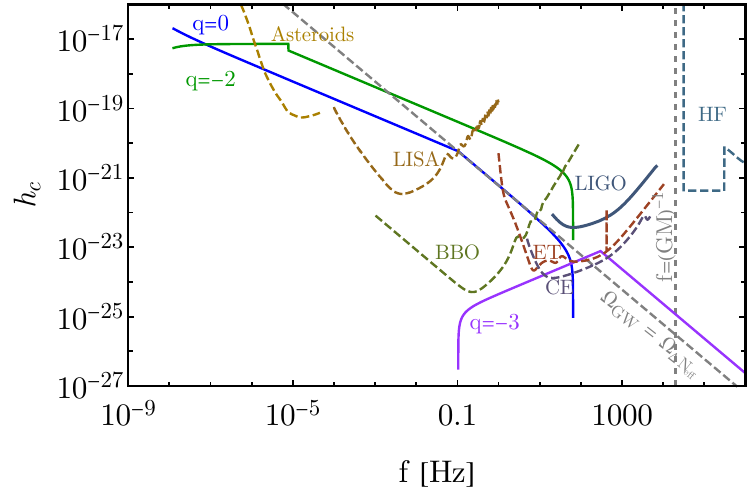}
\end{minipage}
\vskip\baselineskip

\begin{minipage}{.48\textwidth}
    \centering
    \includegraphics[width=0.995\linewidth]{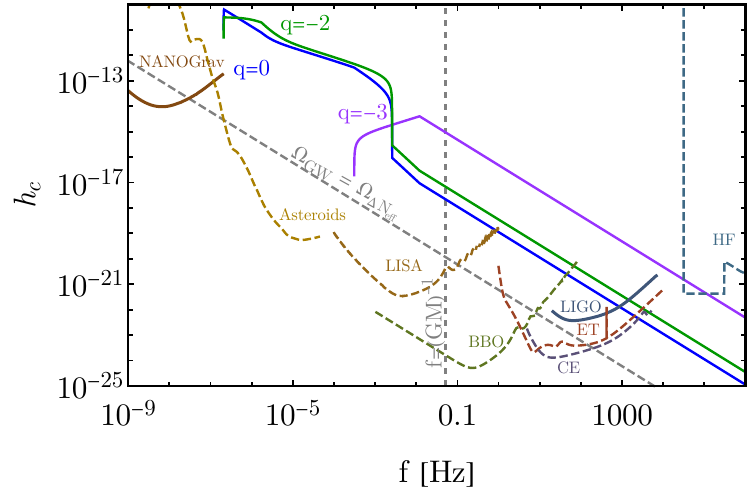}

\end{minipage}%
\hfill
\begin{minipage}{.48\textwidth}
    \centering
    \includegraphics[width=0.995\linewidth]{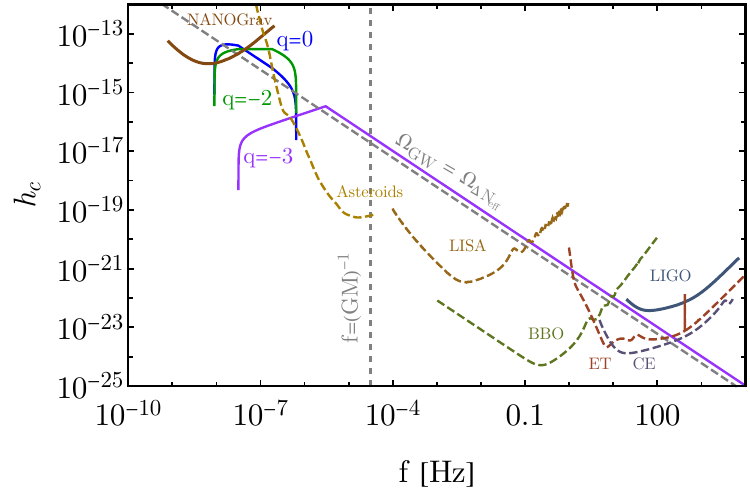}
\end{minipage}
\caption{An example of the strain spectrum generated during the decoupled regime for parameter choices the same as in Figure~\ref{fig: tightly coupled spectrum example}. The strain spectrum in the decoupled regime is low frequency dominated for almost all choices of the spectral index $q$, and this regime starts roughly $3 t_{\rm trans}$ after the string bosenova, lasting until the string network disappears. The signal in the PTA frequency range are stochastic in nature, and are likely already constrained by data. If the duration of the signal is longer than $20$ years, we include only the first $20$ years of the signal.    }
    \label{fig: decoupled spectrum example}
\end{figure*}

To get the characteristic strain in the IR dominated regime ($q>-1$), we follow similar steps as in the tightly coupled regime. The emission comes from both the period where the string-string interaction is efficient, where\footnote{Here we neglected the strings with length smaller than $l_{\rm min}$ that has decoupled from the string-string interaction. In principle, they contribute to a small UV tail which is observationally not relevant.} 
\bea \label{eq: hc DPdecoupled IR}
h_{c, {\rm ssc}}(f) \approx \frac{ \sqrt{2 \Gamma_{\rm GW} }L_{\rm tot}  G \mu}{ \pi d } \cdot
\begin{cases}  
 \l \frac{ \Lt}{L_{\rm tot}} \r^{\frac{1}{2}} \l \frac{f t_{\rm trans}}{2} \r^{\frac{-2-q}{2}}  \quad  & 2 m_{A'} > f > \frac{2}{\Lt}
\\
\l \frac{f L_{\rm tot}}{2} \r^{-\frac{1}{2}}  \quad & \frac{2}{\Lt} > f > \frac{2}{L_{\rm tot}} 
\\
0  \quad & f > 2 m_{A'} \text{ or } f < \frac{2}{L_{\rm tot}} 
\end{cases} .
\eea
as well as the later period when the $l \sim L_{\rm tot}$ string decays through dark photon emission giving
\bea \label{eq: hc decoupled IR}
h_{c, {\rm dec}}(f) \approx \frac{ \sqrt{( \Gamma_{\rm GW}/\Gamma_{\rm DP}) L_{\rm tot}^3 \mA}  G \mu }{ \pi d }  \cdot \begin{cases}  
  \l \frac{f L_{\rm tot}}{2} \r^{-3/2}  & 2 m_{A'} > f > \frac{2}{L_{\rm tot}} \\
\l m_{A'} L_{\rm tot}  \r^{-1/2} \l \frac{f L_{\rm tot}}{2} \r^{-1}  & f > 2 m_{A'}
\\
0 & f < \frac{2}{L_{\rm tot}}
\end{cases}  .
\eea
Comparing equation~\ref{eq: hc DPdecoupled IR} and~\ref{eq: hc decoupled IR}, the latter is larger by a factor of $\sqrt{L_{\rm tot} \mA}$, coming from the ratio between the lifetime of the string of length ($\mA L_{\rm tot}^2$) and the duration of the period where the string-string interaction is efficient ($L_{\rm tot}$). For some of the relevant parameter space, the former (and maybe even the latter) will be replaced by the duration of the experiment when Figure.~\ref{fig: decoupled spectrum example} is made. 

\paragraph{UV dominated distributions : }

For a network that is UV dominated, a string of length $\sim 1/m_{A'}$ disappears into dark photon radiation before it can interact with other strings almost right after $t_{\rm trans}$, and therefore the strain of the emitted gravitational wave is peaked at frequencies $f \sim m_{A'}$. For strings that are longer than $1/m_{A'}$, they will interact with each other for a while, and decouple as the total length and the density of strings decrease. Such a process is generally much faster than the time scale $t_{\rm trans}$, suggesting that we can treat the volume the string network occupies to be a constant. Analogous to the IR dominated case, we can work out the evolution of the network by solving
\begin{equation}
    \frac{{\rm d} L_{\rm tot} (t)}{{\rm d} t} = -\frac{\Gamma_{\rm DP}L_{\rm tot} (t)}{\mA l_{\rm min}(t)^2}
\end{equation}
with $l_{\rm min}(t) = t_{\rm trans} \l \frac{1}{\mA L_{\rm tot} (t)}\r^{1/3}$ determined by the string-string interaction rate. This allows us to solve for the time evolution of both $L_{\rm tot} (t)$ and $l_{\rm min}(t)$ to be 
\begin{equation}
     \frac{L_{\rm tot} (t)}{L_{\rm tot} (t_{\rm trans})} = \l \frac{1}{1+ \frac{2}{3}\Gamma_{\rm DP} m_{A'} (t-t_{\rm trans})}\r ^{3/2}, \quad l_{\rm min}(t) \simeq \frac{1}{m_{A'}}\l {1+ \frac{2}{3}\Gamma_{\rm DP} m_{A'} (t-t_{\rm trans})}\r ^{1/2},
\end{equation}
 and determine the characteristic strain as
\bea \label{eq: hc decoupled uv}
h_{c, {\rm dec}}(f) \approx \frac{ \sqrt{2 \Gamma_{\rm GW} L_{\rm tot} m_{A'}^{-1}} G \mu }{ \pi d } \cdot \begin{cases}  
  \l \frac{f }{2 m_{A'}} \r^{1/2}  & 2 m_{A'} > f > \frac{2}{\Lt} \\
  \l \frac{f }{2 m_{A'}} \r^{-1}  & f > 2 m_{A'}
\\
0 & f < \frac{2}{\Lt}
\end{cases} 
\eea
in the UV dominated regime ($q<-2$).

The above expressions have been simplified for clarity and include leading contributions to the strain, and is computed in the region of parameter space where the total duration of the signal is shorter than the duration of experiments. Spectra shown in Fig.~\ref{fig: decoupled spectrum example} are obtained from full expressions that can be found in the Appendix~\ref{Appendix: exact results}.

Compared to the tightly coupled regime, where the peak strain is quite independent of $q$, in the decoupled regime, the GW strain is larger in the IR dominated regime and smaller in the UV dominated regime. Physically, this comes from the fact that the fraction of energy dissipated into gravitational wave radiation depends on the ratio between the gravitational wave emission power and the dark photon emission power, which is larger for longer strings. This manifest as a dependence on the duration when the network is emitting at peak frequency $h_c \propto \sqrt{t_{\rm duration}}$, which is $t_{\rm trans}$ in the tightly coupled regime, $L_{\rm tot}$ when the strings-string interaction is efficient and $ \mA L_{\rm tot}^2$ when all interactions are decoupled in the IR dominated case, and $1/m_{A'}$ in the UV dominated case. This dependence is replaced by the duration of the experiment when $t_{\rm duration} > 20 \,{\rm yrs}$, which we take to be the typical duration of GW experiments. The peak strain is
\bea
h_{c,{\rm peak}} &\approx& \frac{ \sqrt{ (\Gamma_{\rm GW}/\Gamma_{\rm DP}) L_{\rm tot}^3 \mA}  G \mu }{ \pi d } \nonumber\\
&\sim&  10^{-19} 
\l \frac{\lambda/g^2}{10^3} \r^{\frac{3}{2}} \l \frac{G\mu}{10^{-14}} \r \l \frac{M_{\rm BH}}{10 M_{\odot}} \r^{-\frac{9}{2}} \l \frac{\mA}{10^{-12} \eV} \r^{-\frac{11}{2}} \l \frac{d}{1 \,\text{kpc}} \r^{-1} \l \frac{\Gamma_{\rm GW}}{\Gamma_{\rm DP}} \r^{1/2}. 
\eea
for the IR dominated case ($q>-1$), and
\bea
h_{c,{\rm peak}} &\approx& \frac{ \sqrt{2 \Gamma_{\rm GW} L_{\rm tot} \mA^{-1}}  G \mu }{ \pi d } \nonumber\\
&\sim&  10^{-21} 
\l \frac{\lambda/g^2}{10^5} \r^{\frac{1}{2}} \l \frac{G\mu}{10^{-14}} \r \l \frac{M_{\rm BH}}{10 M_{\odot}} \r^{-\frac{3}{2}} \l \frac{\mA}{10^{-13} \eV} \r^{-\frac{5}{2}} \l \frac{d}{1 \,\text{kpc}} \r^{-1} . 
\eea
for the UV dominated case ($q<-2$).

\subsection{Overall signal}
Now, we have all the necessary ingredients to discuss the overall GW signal that reaches the earth. 
At any given stage, the probability that there is a current ongoing event given by its duration, $\Delta t N_B$, divided by the average formation time of black holes in our galaxy.
\bea
P_{\rm sig} = R_\text{BHF} \, N_B \, \Delta t
\eea
where $N_B = \frac{\alpha^5}{(\lambda/g^2) G\mu}$ is the number of bursts in the lifetime of the BH, and $R_\text{BHF}$ is the Black Hole Formation rate in the Milky Way.  When numerical input is required, we will take $R_\text{BHF} = 0.9$ per century \cite{Baryakhtar:2017ngi}.
When $P_{\rm sig}$ approaches 1, then we start seeing overlapping signals from multiple bosenova events.   Somewhat unfortunately, the different parts of the spectrum have different duration.  The lowest frequencies can easily have a duration that is parametrically longer than years, while the higher frequencies can last less than a second.  For the parameters chosen in the figures discussed below, $P_{\rm sig} < 1$ for all but the very lowest of frequencies.

The GW signal that reaches the earth is a combination of a tightly coupled regime and the decoupled regime. The initial signal consists of only the tightly coupled regime and after $\sim \Lt$, the signal from the closest strings that decoupled arrives. As the cloud of strings expands in all directions with $v \sim c$, signals from closer and further parts of the cloud overlap, mixing the GW signal from tightly coupled and decoupled regimes for another $\sim 2 \Lt$. After this time the remaining signal comes from the decoupled regime only and lasts $\sim 3 \tau$. In the parameter space that we show below in Fig. \ref{fig: Parameter space GW} and Fig. \ref{fig: Parameter space GW combined}, these timescales take values between $ 10^{-3} \, {\rm s} \lesssim \Lt \lesssim 1 \, {\rm hr} $, and $ 1 \, {\rm s} \lesssim \tau \lesssim 10^{20} \, {\rm yr}$. 
The overall characteristic strain after integrating over the duration of signal is given by

\bea \label{Eq: Combined hc}
h_{c,{\rm tot}}(f) = \sqrt{h_{c,{\rm tc}}^2(f) + h_{c,{\rm ssc}}^2(f) + h_{c,{\rm dec}}^2(f) }.
\eea

Examples of spectra $h_{c,{\rm tot}}(f)$ are shown in Fig. \ref{fig: combined spectrum example}. Additionally, Fig. \ref{fig: spectrum contributions example} in the Appendix \ref{Appendix: exact results} compares how $h_{c,{\rm tc}}(f)$, $ h_{c,{\rm ssc}}(f) $, and $ h_{c,{\rm dec}}(f)$ contribute to the signal.  To assess their detectability by current and proposed GW detectors, for each set of parameters we calculate the ${\rm SNR}$ averaged over the distribution of BH in the Milky Way\footnote{We assume that matched filtering has been used when calculating the ${\rm SNR}$. Analysis of real data would require dedicated numerical simulations to produce exact waveforms, especially in the tightly coupled regime.}

\bea\label{Eq: SnR GW}
\bigg< \frac{S}{N} \bigg> = \bigg< \sqrt{\int_{-\infty}^\infty d\log f \frac{\left|h_{c,{\rm tot}}(f) \right|^2}{ \left| h_n(f) \right|^2}} \bigg>_{M_{\rm BH},d}.
\eea
Plots in Fig. \ref{fig: Parameter space GW} and Fig. \ref{fig: Parameter space GW combined} highlight parameters accessible by current and future GW experiments assuming $q=0$.
The shaded regions in Fig. \ref{fig: Parameter space GW} show parameter space where the expected signal-to-noise ratio is ${\rm SNR} > 5$ for a given experiment (assuming $q=0$). 
Purple represents LIGO-VIRGO-KAGRA \cite{KAGRA:2021tnv}, pink LISA \cite{Colpi:2024xhw}, green BBO \cite{Cutler:2009qv}, red ET \cite{Hild:2009ns}, and dark blue CE \cite{Srivastava:2022slt}. ${\rm SNR} < 5$ below the colored lines while the upper part of the shaded region is cut-off when the rate of bursts modulo 1-sigma fluctuations reaches one every 10 years.  
We require that the rate up to 1-sigma fluctuations is at least
$\Gamma_{\rm MW} + \sigma_{\Gamma_{\rm MW}} > (10 \text{ yr})^{-1}$. The value of $\lambda/g^2$ is fixed to $\lambda/g^2 = 10^4$ (left) and $\lambda/g^2 = 10^8$ (right). Additionally, gray lines provide a reference on the average rate of bursts. 
To the right of the middle line, bursts would occur at least once a day, and on the right of the top line, they would occur at least once a decade.
Each of these lines ends when there is not enough active BH in the Milky Way to produce the signal at the given rate. The cutoff on the right is due to not satisfying the condition for superradiance.
Note that the signal in the decoupled regime from a single long string disappearing via dark photon and gravitational wave emission can in principle last for longer than the experimental duration, as well as inverse rate of bursts, and could, as a result, be stochastic in nature. However, the initial tightly coupled and string network expansion phase lasts for a duration that is small compared to the inverse rate of bursts. These short lived regimes provide smoking gun evidences for the origin of the signal being from a string bosenova.

The plot in Fig. \ref{fig: Parameter space GW combined} merges parameter space plots with $\lambda/g^2 = 10^2$, $\lambda/g^2 = 10^4$, $\lambda/g^2 = 10^6$, and $\lambda/g^2 = 10^8$. Each value of $\lambda/g^2$ is represented by a different color, with the darker shade covering the parameter space accessible by LIGO and the lighter shade covering the parameter space that would be probed by any of the proposed experiments that appear in Fig. \ref{fig: Parameter space GW}. The bands with $\lambda/g^2 = 10^2$ start at higher values of $G\mu$, while bands with higher values of $\lambda/g^2$ start lower respectively.

\begin{figure*}[t] 
\centering
\begin{minipage}{.48\textwidth}
    \centering
    \includegraphics[width=0.995\linewidth]{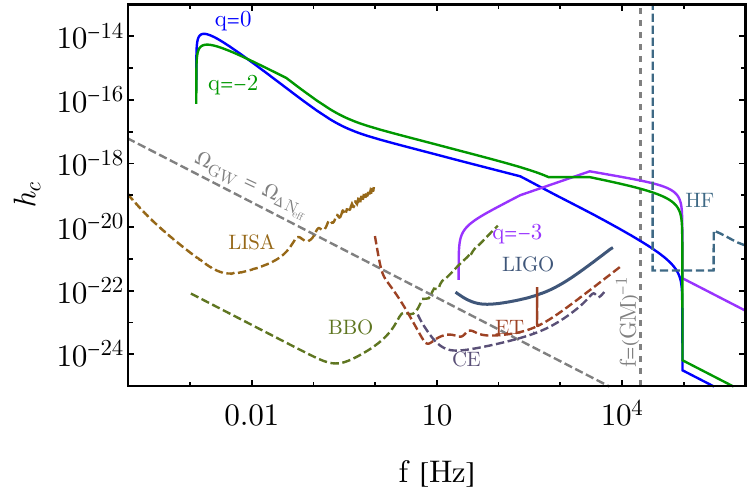}
\end{minipage}%
\hfill
\begin{minipage}{.48\textwidth}
    \centering
    \includegraphics[width=0.995\linewidth]{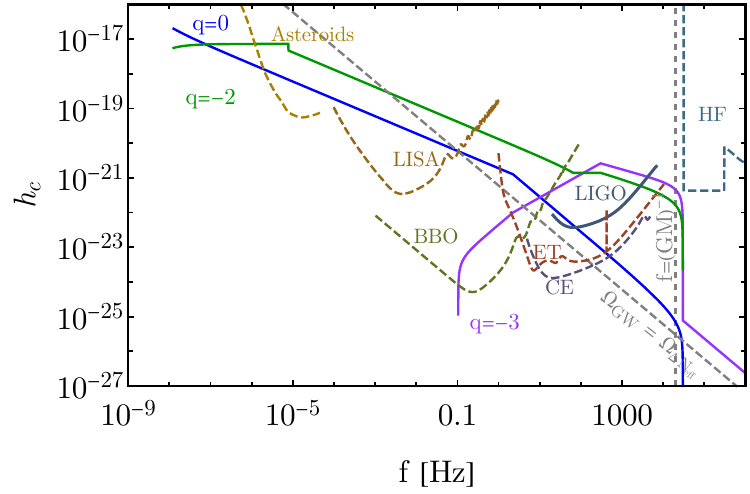}
\end{minipage}
\vskip\baselineskip

\begin{minipage}{.48\textwidth}
    \centering
    \includegraphics[width=0.995\linewidth]{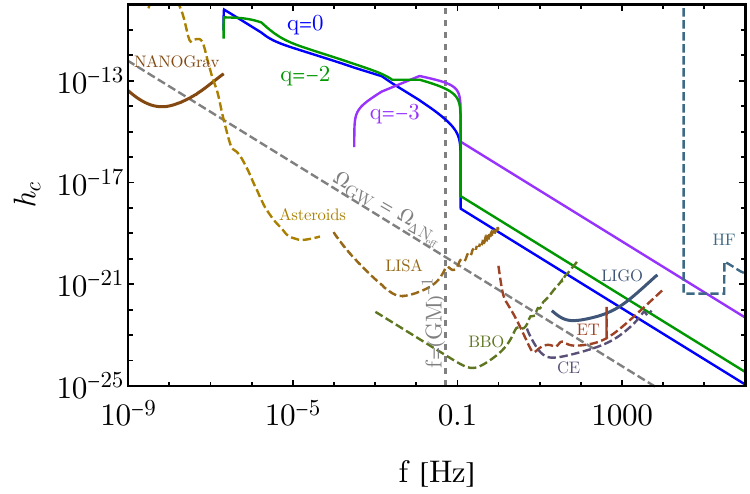}

\end{minipage}%
\hfill
\begin{minipage}{.48\textwidth}
    \centering
    \includegraphics[width=0.995\linewidth]{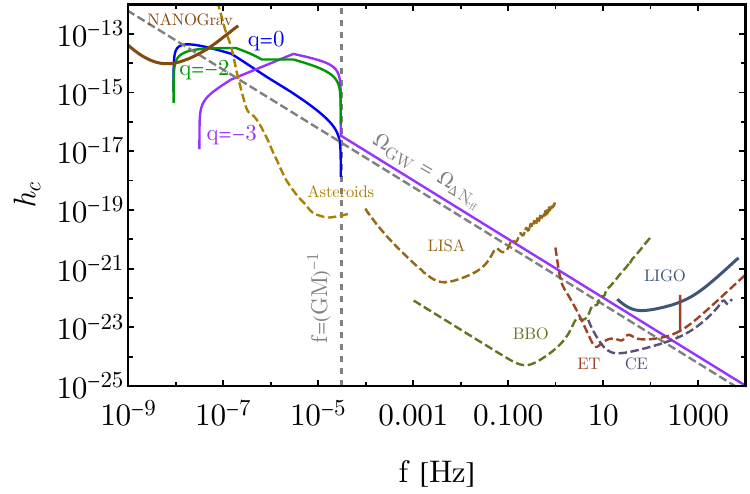}
\end{minipage}
\caption{The average strain spectrum generated during the lifetime of the whole string network for parameter choices the same as in Figure~\ref{fig: tightly coupled spectrum example} and different spectral index $q$. Note that this is different from the strain spectrum in the period between $\sim t_{\rm trans}$ and $\sim 3 t_{\rm trans}$ after a string bosenova event, which will be a simple sum of the strain in Figure~\ref{fig: tightly coupled spectrum example} and~\ref{fig: decoupled spectrum example}. If the duration of the signal is longer than $20$ years, we include only the first $20$ years of the signal.  
}
    \label{fig: combined spectrum example}
\end{figure*}

\begin{figure*}[t] 
\centering
\begin{minipage}{.48\textwidth}
    \centering
    \includegraphics[width=0.995\linewidth]{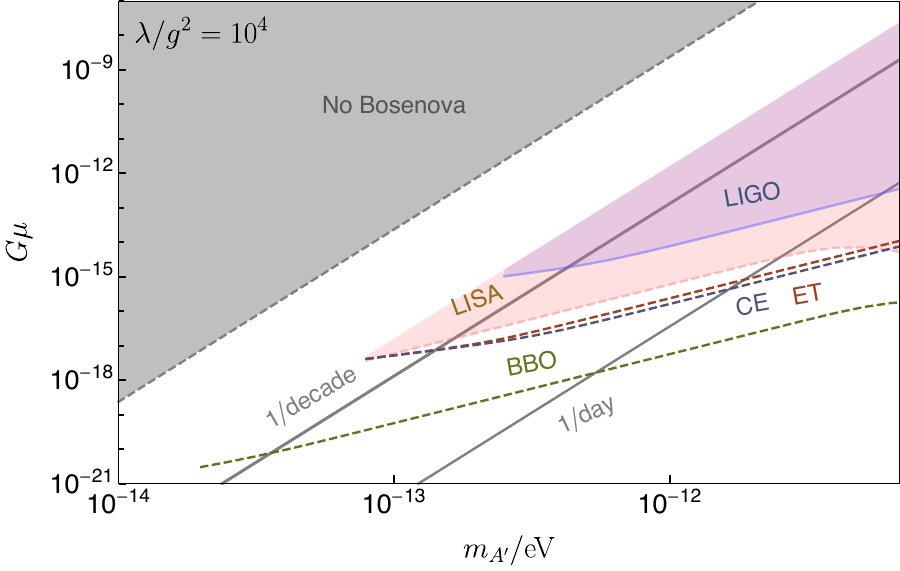}
\end{minipage}%
\hfill
\begin{minipage}{.48\textwidth}
    \centering
    \includegraphics[width=0.995\linewidth]{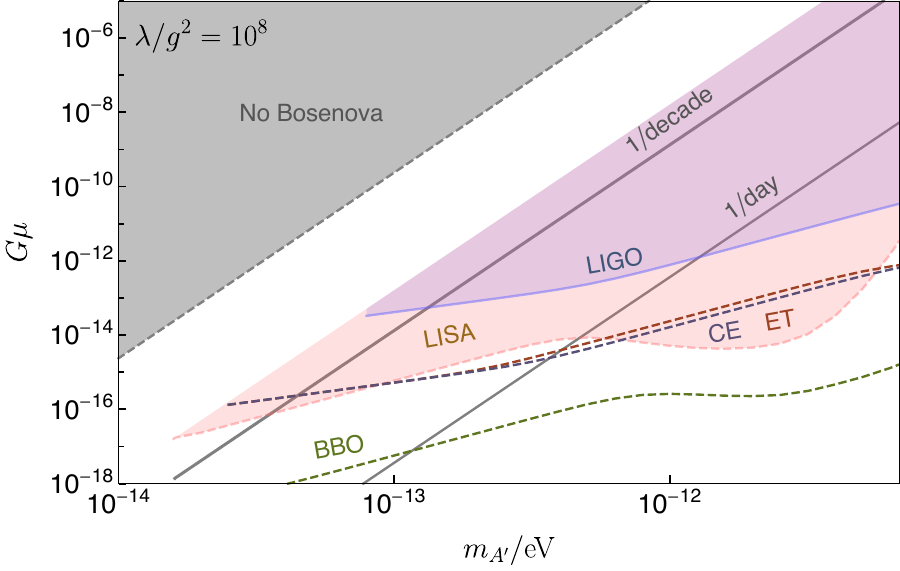}
\end{minipage}

\caption{The regions of $G\mu$ vs. $\mA$ parameter space for $q=0$ where ${\rm SNR} > 5$ can be reached with a dedicated analysis.  The colored bands indicate the parameter space explored by various different experiments.  The solid lines indicate when the event rate is one per day and one per decade. The left plot assumes $\lambda/g^2 = 10^4$, and the right plot assumes $\lambda/g^2 = 10^8$. The signal duration can easily be longer than decades in the parameter region above the colored shaded region. The gravitational wave signal in this region will be stochastic in nature, and would not be smoking gun evidences for the string bosenova. See text for more details.}
    \label{fig: Parameter space GW}
\end{figure*}

\begin{figure*}[t] 

     \centering
     \includegraphics[width=0.8\linewidth]{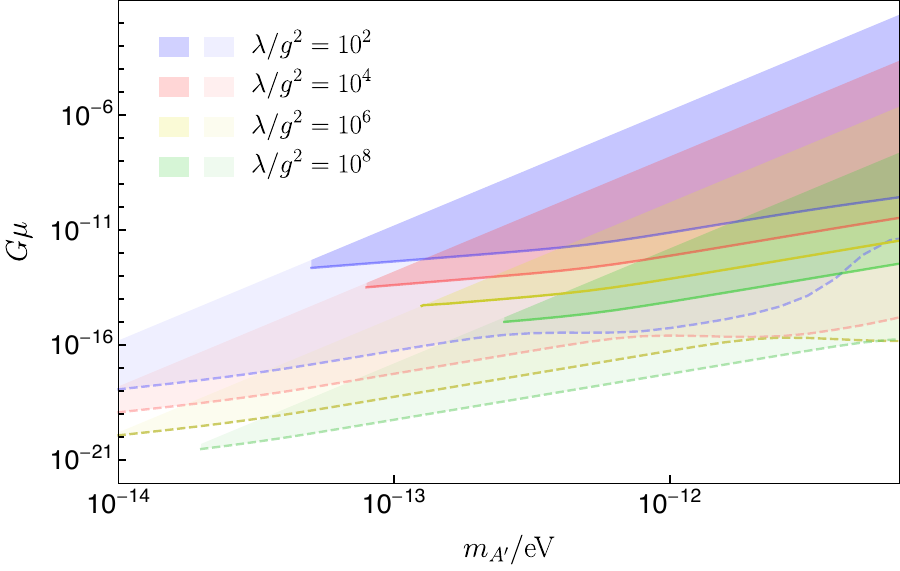}
    
\caption{The $G\mu$ vs. $\mA$ parameter space for $q = 0$ where ${\rm SNR} > 5$ can be reached with a dedicated analysis.  The solid shaded region shows when ${\rm SNR} > 5$ for current ground based detectors (LIGO) for various values of $\lambda/g^2$, while the light shaded region is when ${\rm SNR} > 5$ for future detectors (BBO).}
    \label{fig: Parameter space GW combined}
\end{figure*}

\pagebreak

\section{Direct Detection of B-L Strings}~\label{Sec:DirectImpact}

In this section, we will examine the possibility of directly detecting strings passing through Earth. As discussed in Section ~\ref{Sec: String Dynamics}, in the small $G\mu$, large $\lambda/g^2$ limit, these strings can live long enough to traverse the whole Milky Way and reach the Earth, and interact with Earth based gravitational wave detectors and accelerometers. To illustrate how the strings interact with the standard model particles comprising a detector, we will consider $U(1)_{\rm B-L}$ dark photons as an example of the many different ways that standard model particles can be charged under an $U(1)$ extension to the standard model. In the dark photon mass range we are interested in ($m_{A'}\sim 10^{-15}-10^{-12} \,{\rm eV}$), equivalence principle tests place the strongest constraint on the $B-L$ coupling as $g_{\rm B-L}\leq 10^{-23}-10^{-24}$~\cite{HEECK2014256} which in turn place a lower bound on $G\mu$
\bea
\label{Eq: Gmu Constraints}
G\mu\sim \frac{Gm_{A'}^2}{g_{\rm B-L}^2}\geq 10^{-31}-10^{-37}
\eea
For reasonable $\lambda$, we can easily work in the small $G\mu$, large $\lambda/g^2$ limit. To determine whether or not strings can be detected on Earth, we must compute both the strength of the single produced by a passing string as well as the expected rate at which strings arrive on Earth. To compute the strength of the signal, we start by computing the effect of a single string on a single $B-L$ charged test mass and showing that this effect is independent of the coupling $g_{\rm B-L}$. We then take this single test mass result and use it to compute the signal-to-noise ratio (${\rm SNR}$) of a single string on various gravitational wave detectors and other accelerometers. We then compute the expected rate of strings passing through Earth to determine in which region of parameter space strings pass frequently enough to be detected by these experiments.

\subsection{String Passing a Single Mass}
\begin{figure}[h!]
\centering
\includegraphics[width=.7\linewidth]{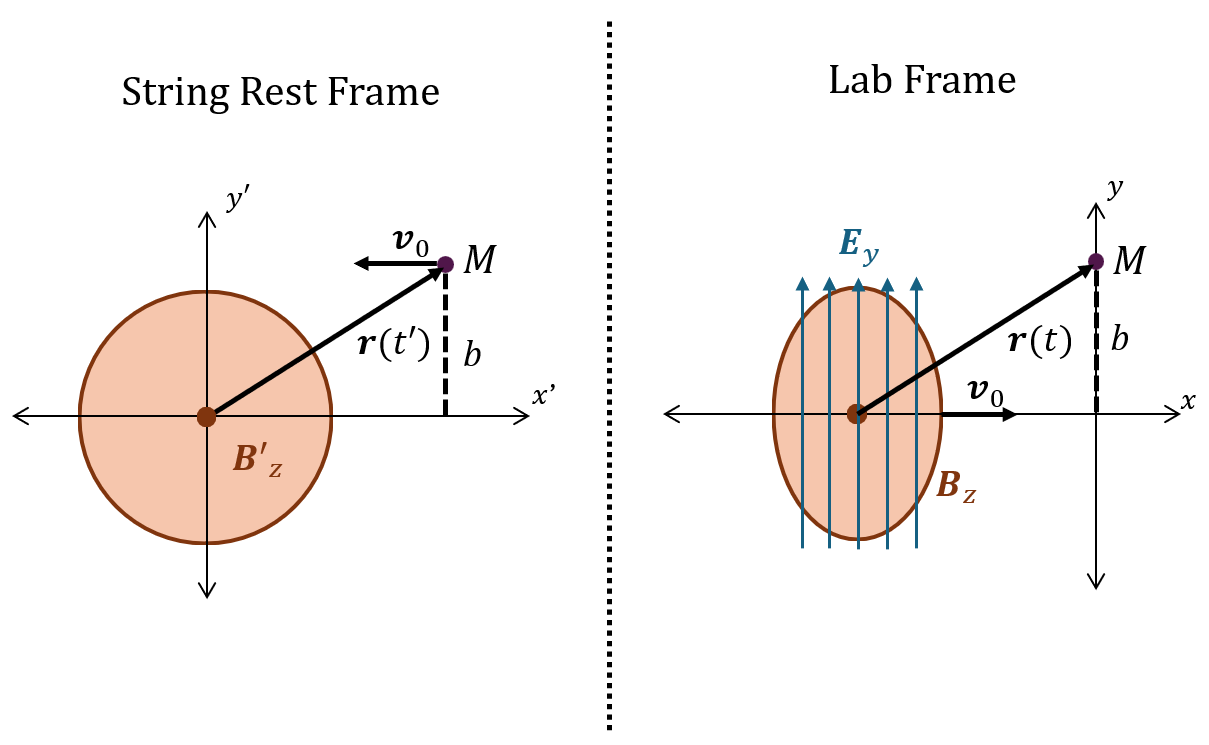}
\caption{The collision of a string and a test mass $M$ as seen in both the strings rest frame and the initial rest frame of the test mass (lab frame). In both, the string's magnetic field points in the positive $z$-direction out of the page. } 
\label{Fig: Single Mass}
\end{figure}

Consider a single string with velocity $v_0$ passing by a single test mass $M$ with ${ B-L}$ charge $Q_{\rm B-L}$. For simplicity, we orient our coordinates so that the string is oriented with its magnetic field in the positive $z$-direction and its velocity in the positive $x$-direction. A sketch of this scenario is shown in figure~\ref{Fig: Single Mass} in both the string's rest frame and the initial rest frame of the test mass. Note the ${B-L}$ magnetic field of the dark photon string changes over $\gtrsim \,{\rm km}$ length scales, ensuring that we can always treat the test masses as point particles. 

As computed by~\cite{East:2022rsi}, in its rest frame, the string has a radial magnetic field profile
\bea
\label{Eq: String Magnetic Field}
B'_z= \frac{m_{A'}^2}{g_{\rm B-L}}K_0(m_{A'} r)
\eea
Where $K_0$ is the zeroth modified Bessel function of the second kind. It should be noted that the magnetic flux through a radial cross-section of the string is a single flux quantum, $2 \pi/g_{\rm B-L}$. Boosting to the rest frame of the test mass, the string acquires an electric field $E_y= \gamma v_0 B'_z$ perpendicular to its magnetic field and its velocity. In the limit where the test mass is nonrelativistic in the lab frame, this electric field gives the dominant contribution to the Lorentz force
\bea
\label{Eq: Coordinate Lorentz Force}
\frac{d \vec p}{dt}=g_{\rm B-L}Q_{\rm B-L}\vec E(r(t'))=g_{\rm B-L}Q_{\rm B-L}\gamma v_0\frac{m_{A'}^2}{g_{\rm B-L}}K_0\l m_{A'} \sqrt{b^2+(\gamma v_0 t)^2}\r\hat {\vec{y}}
\eea
Where $b$, as shown in figure~\ref{Fig: Single Mass}, is the impact parameter of the mass with respect to the center of the string. Immediately, we can see that since the $g_{\rm B-L}$ from the charge of our test mass cancels exactly with the $g_{\rm B-L}^{-1}$ from the strings electric field, this force is totally independent of the coupling! 
In general, we find
\bea
\label{Eq: Lorentz force}
\frac{d^2 \vec r(t)}{dt^2}=\frac{\gamma v_0 m_{A'}^2}{2m_p}K_0\l m_{A'} \sqrt{b^2+(\gamma v_0 t)^2}\r( \hat {\vec B}\times \hat {\vec{v_0}})
\eea
Where $\hat{\vec v
}_0$ and $\hat{\vec B}$ are unit vectors pointing along the string's velocity and the string's magnetic field respectively. Here we assume that the test mass contains an equal number of protons and neutrons so that $Q_{\rm B-L}/M=g_{\rm B-L}/2m_p$
We can solve this equation by Fourier transforming both sides to get
\bea
\label{Eq: Displacement solution}
\tilde{\vec r}(f)=-\frac{m_{A'}^2 e^{-b\Omega(f)}} {16 \pi m_pf^2\Omega(f)}( \hat {\vec B}\times \hat {\vec{v_0}})
\eea
where  
\bea
\label{Eq: Fourier Transf Def}
\Omega(f)=\sqrt{m_{A'}^2+\l\frac{2 \pi f}{\gamma v_0}\r^2}\quad \text{and} \quad \tilde{\vec r}(f)=\int dt e^{2\pi i f t}\vec r(t)
\eea

\subsection{Signal At Gravitational Wave Detectors}

Strings leave a signal when they pass through accelerometers, in particular, various current and future gravitational wave detectors. Generally, each gravitational wave detector consists of some configuration of test masses and the experiment is sensitive to strains in the configuration. While normally the strain in the test masses is caused by gravitational waves, here we consider the possibility of the strain being caused by a passing string, with a characteristic frequency spectrum. Using Eq.~\ref{Eq: Displacement solution}-~\ref{Eq: Fourier Transf Def}, we can compute the displacements of the test masses in the gravitational wave detector due to the passing string and thus determine the signal measured by the detector. We then compute the ${\rm SNR}$ for these detectors to determine if the passing string is detectable or not. We consider both current and future gravitational wave detectors by looking at LIGO~\cite{LIGOScientific:2014pky}, MAGIS~\cite{Coleman_MAGIS100} and the Einstien Telescope~\cite{Punturo:2010zz}. We begin with a brief description of each detector.\\

\noindent \textbf{\uline{LIGO}}\\
\indent The Laser Interferometer Gravitational-Wave Observatory (LIGO) are specialized Michaelson interferometers with $4$ km long arms. The sensitivity of the interferometer in enhanced by two Fabry-Perot cavities (one in each arm) each with length $L_0=4$ km. Both cavities have mirrors on either end which play the role of test masses in the gravitational wave detector. LIGO is sensitive to the relative strain between the two cavities, $h$, which is defined as 
\bea
\label{Eq: LIGO strain}
h(t)=\frac{L_x(t)-L_y(t)}{L_0}
\eea
Where $L_x$ and $L_y$ are the lengths of the two cavities. In particular, LIGO is sensitive to the Fourier transform of this strain $\tilde h(f)$ for frequencies between roughly $10$ Hz and $1$ kHz. The sensitivity curve can be found in Figure 2 of~\cite{LIGOScientific:2014pky}.\\
\noindent \textbf{\uline{MAGIS-100}}\\
\indent The Matter-wave Atomic Gradiometer Interferometric Sensor (MAGIS-100), is a future gravitational wave detector that aims to detect gravitational waves and other signals of new physics via a 100 meter baseline atomic interferometer at Fermilab. The experiment will consist of cold clouds of atoms in free fall upon which three successive laser pulses act as the mirrors and beam splitters of a Mach-Zender interferometer. MAGIS-100 projects to measure the strain $h(f)$ between these two clouds of atoms with a sensitivity of $10^{-21}$ Hz$^{-1/2}$ in the $0.1-10$ Hz frequency range as shown in figure 1 of~\cite{Coleman_MAGIS100}.\\ 
\noindent \textbf{\uline{Einstein Telescope}}\\
\indent The Einstein Telescope (ET) is a future gravitational wave detector based of the same Fabry-Perot enhanced Michaelson interferometer technology as LIGO. Unlike LIGO, ET will consist of 3 interferometers, each with an angle of $60^\circ$ between their $10$ km length arms. The 3 interferometers are arranged to form an equilateral triangle as shown in~\cite{Rowlinson_2021}. They expect to measure the strain with a sensitivity roughly an order of magnitude better than LIGO with the sensitivity curve shown in figure 2 of~\cite{Punturo:2010zz}.\\

For each of these experiments, we can compute the signal-to-noise ratio of a passing string as 
\bea\label{Eq: Straing SnR}
\frac{S}{N}=\sqrt{\int_0^\infty df \left|\frac{h(f)}{S_n^{1/2}(f)}\right|^2 }
\eea
In each experiment, the strain produced by the string will depend on the orientation of the string relative to the detector, the impact parameter between the center of the string and the detector, and the velocity of the string. To account for this, we simulate 10,000 collision events each with a random orientation, a random impact parameter ranging from $0$ to $1/m_{A'}$, and a random velocity ranging from $.01$ to $.99$. In figure~\ref{Fig: SNR} we plot the average ${\rm SNR}$ for each experiment as a function of $m_{A'}$ with the $99\%$ confidence region shown in the shaded regions.

\subsection{Signal At EP Violation Tests}

The gravitational wave detectors are sensitive to forces/acceleration from passing strings independent of whether this force violates the equivalence principle. The signal, therefore, is gradient suppressed (spatial and temporal) in the limit of zero dark photon masses. Composition dependent equivalence principle violation tests have been used to achieve better sensitivity in searches for light bosons at low masses~\cite{Arvanitaki:2014faa,Shaw:2021gnp}. The same would happen in searches for passing of topological defects~\cite{Pospelov:2012mt} if they accelerate masses in a weak equivalence principle violating manner. In this subsection, we use the Windchime project~\cite{Windchime:2022whs} as an example to demonstrate this point.

\begin{figure}[h!]
\centering
\includegraphics[width=.80\linewidth]{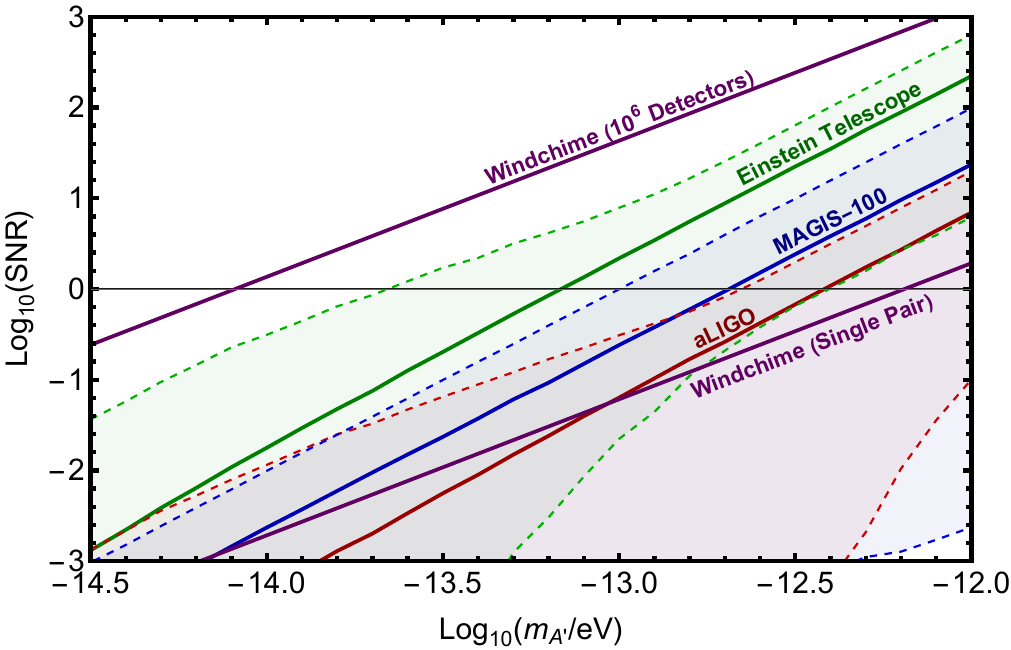}
\caption{A plot of the signal-to-noise ratio per string of a passing dark photon string at various gravitational wave detectors. The solid lines represent the average ${\rm SNR}$s while the shaded region represents the $99\%$ confidence region for the ${\rm SNR}$ at a given detector based on simulated events with random string orientation relative to the detector, random string velocity between $.01$ and $.99$, and random impact parameter $b$ between 0 and $1/m_{A'}$. We also show the ${\rm SNR}$ for a string passing by Windchime an array of quantum-enhanced accelerometers with sensitivity $S_n^{1/2}=10^{-13}$ m/s$^2$/$\sqrt{\text{Hz}}$, and $\Delta=0.05$. We have averaged the ${\rm SNR}$ of windchime over impact parameters $b$ and velocities, ranging from $0$ to $1/m_{A'}$ and $.01$ to $.99$ respectively. We show the ${\rm SNR}$ for a single pair of detectors and for an array of $10^6$ detectors.} 
\label{Fig: SNR}
\end{figure}

Windchime is a proposed future dark matter detector aiming to detect Planck-scale mass dark matter directly through its gravitation interaction as well as detect certain models for ultra-light dark matter that would be capable of detecting passing B-L strings. Windchime consists of an array of quantum-enhanced mechanical accelerometers capable of tracking and detecting their acceleration due to nearby objects. A passing B-L string would produce an acceleration on such sensors, as can be seen from Eq.~\ref{Eq: Displacement solution}, and are therefore prime candidates to be detected by Windchime. From Eq.~\ref{Eq: Displacement solution}-\ref{Eq: Fourier Transf Def}, we can easily compute the Fourier transform of the relative acceleration, $\tilde a(f)$, between two sensors with differing neutron to ratio. If $N_{1,2}$ and $A_{1,2}$ are the number of neutrons and the atomic mass number of two neighboring sensors in the array and define $\Delta=\frac{N_1}{A_1}-\frac{N_2}{A_2}$, we find the relative acceleration to be 
\bea\label{Eq: relative acceleration}
\tilde a(f)=\Delta\frac{\pi m_{A'}^2e^{-b\Omega(f)}}{2m_p\Omega(f)}
\eea
Windchime expects to be able to reach sensitivities in acceleration of around $S_n^{1/2}=10^{-13}$ m/s$^2$/$\sqrt{\text{Hz}}$ over a wide range of frequencies~\cite{Windchime:2022whs}. We can therefore easily compute the signal-to-noise ratio for a pair of their sensors for a passing string
\bea
\label{Eq: SNR Windchime}
\frac{S}{N}=\sqrt{\int_0^\infty df \left| \frac{a(f)}{S_n^{1/2}}\right|^2}=\frac{ I(bm_{A'})\Delta}{S_n^{1/2}}\sqrt{\frac{\pi\gamma v_0m_{A'}^3}{8m_p^2}} 
\quad \text{where,}\quad I(a)=\sqrt{\int_0^\infty dx\; \frac{e^{-2a\sqrt{1+x^2}}}{1+x^2}}
\eea
For $m_{A'}b \sim \OO(1)$, $I(m_{A'}b)$ is some order one factor. If we then assume $v_0\gamma\sim m_{A'}b\sim \OO(1)$ we can estimate the ${\rm SNR}$ as 
\bea
\label{Eq: SNR Windchime Estimate}
\frac{S}{N}\approx 5 \l\frac{\Delta}{0.05}\r\l\frac{m_{A'}}{10^{-12} \text{eV}}\r^{3/2}\l\frac{S_n^{1/2}}{10^{-13}\text{m}/\text{s}^2/\sqrt{\text{Hz}}}\r^{-1}
\eea
It is important to note that this ${\rm SNR}$ gets enhanced by a factor of $\sqrt{N_{\rm det}}$ where $N_{\rm det}$ is the number of pairs of detectors in the array. If we take an array of $10^6$ detectors, this would allow Windchime to be sensitive to passing strings down to $m_{A'}\approx 5\cdot 10^{-15}$ eV. Figure~\ref{Fig: SNR} shows the ${\rm SNR}$ of Windchime with an array of $10^6$ detectors with sensitivity $S_n^{1/2}=10^{-13}$ m/s$^2$/$\sqrt{\text{Hz}}$, $\Delta=0.05$, averaged over impact parameters $b$, ranging from $0$ to $1/m_{A'}$, and averaged over velocities $v_0$ from $.01$ to $.99$.

\subsection{Rate of String Arrival on Earth}
Of equal importance as their signal in gravitational wave detectors is the frequency with which strings collide with these detectors. If the rate of string arrivals, $\Gamma_{\rm string}$, is $\gtrsim 1$ yr$^{-1}$ we can be confident that any of the aforementioned detectors will experience one if not multiple collisions in their experimental run time.
Here we will compute the rate $\Gamma_{\rm strings}$, first by considering a singular string burst from a single black hole, before building up to multiple bursts from a single black hole, and finally considering the full ensemble of all black holes in our galaxy to get the full, true rate of string arrival.

To start let us find $\Gamma_{\rm strings}$ from a single burst from a single black hole of mass $M_{\rm BH}$ a distance $d$ from Earth. As discussed in section~\ref{Sec: String Dynamics}, after the burst the string network scales until it reaches a critical size $t_{\rm trans}$ given in Eq.~\ref{Eq: ttrans}. After this strong regime, the strings free stream across the galaxy. One can then easily compute the flux of strings on earth with initial length $l_0$ at decoupling, $\frac{d\Phi}{dl_0}$, to be
\bea
\label{Eq: Flux of Earth}
\frac{d\Phi}{dl_0}=\frac{1}{4\pi d^2 t}F_s(l_0,t_{\rm trans})
\eea
Where $F_s(l_0,t_{\rm trans})$ is the number of strings per length $l_0$ at the transition to the free streaming regime. As the strings travel, they decrease in length due to emission into dark photons. By the time they reach earth, strings that started with length $l_0$ will have length $l(l_0,t)$ where $t$ is the time since they entered the free streaming regime. If we assume the size of the string is large compared to the size of the detector, the cross-section of the string and the detector will be roughly the cross-sectional area of the string $A(l)\sim l m_{A'}^{-1}$. The total rate of strings of all lengths arriving on Earth is then
\bea
\Gamma_{\rm strings}=\int dl_0 \frac{d\Phi}{dl_0}A(l(t,l_0))=\frac{1}{4\pi d^2 t m_{A'}}\int dl_0 l(t,l_0)F_s(l_0,t_{\rm trans})
\eea
Upon investigation, one can see that this remaining integral is just $L_{\rm tot}(t)$, the total length of all the strings at time $t$. 
The initial total length of the strings at decoupling, $L_0$, in the $G\mu\ll 1$ limit is the same as the total length after the initial phase transition $L_0 \approx \frac{\lambda}{g^2\alpha^3 m_{A'}}$. The total rate can then be written  
\bea
\label{Eq: Single Burst Rate}
\Gamma_{\rm strings}=\frac{\lambda}{g^2\alpha^3}\frac{\mathcal{F}_L(d)}{4\pi (m_{A'} d)^3}m_{A'}
\eea
Where we have replaced $t=d$ due to the $\OO(1)$ velocity of the string and where $\mathcal{F}_L(t)=L_{\rm tot}(t)/L_0$ is the fraction of the initial total length that remains after time $t$ in the free streaming regime. This is the only part of the rate that depends on the string length distribution $F_s(l,t_{\rm trans})$ and so we can use it to examine the effect of the shape of the distribution on $F_s$ on the rate. In appendix~\ref{Apdx: FL computation} we give both a numerical calculation and an analytic approximation of $\mathcal{F}_L$ for power law distributions $F_s(l)\propto l^{q}$. We find that for $q<-2$, $\mathcal{F}_{L}(d)\ll 1$  for distances $d\sim \OO(kpc)$, due to the majority of the length being concentrated in strings of minimal length which decay very quickly. We also find that for $q>-2$, after averaging the rate over distributions of black hole mass and distance, the rate is independent of $q$. As such, going forward in this section we will consider only distributions with $q>-2$ for which the results are equal. \\\\
As discussed in section~\ref{Sec: String Dynamics}, the black hole repeats its Bosenova process until all the available black hole spin has been depleted, a total of $\mathcal{F}_s^{-1}$ string bursts. Each of these Bosenovae is separated by a time $t_{\rm SH}$ given in Eq~\ref{Eq: Superheating Time}. From Eq.~\ref{Eq: Superheating Time}, we can see that $t_{\rm SH}\propto m_{A'}^{-7}$ for a fixed black hole mass and thus can span numerous orders of magnitude. This results in behaviors ranging from Bosenovae happening in quick secession to being Bosenovae separated by millennia. From our point of view on earth, these Bosenovae last for a time $d\sim 10^5$ yr due to the $\mathcal{O}(1)$ in the strings velocity dispersion. If $t_{\rm SH}$ is long, the Bosenova bursts from the black hole would be separated by periods where there were no strings arriving. On the other hand, if $t_{\rm SH}$ is small, we expect Bosenova bursts to overlap here on Earth, leading to an enhancement of the overall string rate from that of a single burst. We can quantify this by computing $N_{\rm exp}$ as the number Bosenova bursts we would expect to observe on average at any given time. Intuitively, $N_{\rm exp}$ will be the ratio of the total time the black hole spends in its "bursting state" to the total time it spends in the superradiant process $\tau_{\rm BH}$. The total time spent in the bursting state from our point of view on earth is simply the duration of each burst, $d$, times the number of bursts $\mathcal{F}_s^{-1}$
\bea
\label{Eq: Nexp}
N_{\rm exp}=\frac{d}{\mathcal{F}_s\tau_{\rm BH}}
\eea
The rate of strings then is enhanced (or reduced) by this factor. Using Eq.~\ref{Eq: Single Burst Rate} after some simplification, the string arrival rate becomes
\bea
\label{Eq: Multiple Bursts Rate}
\Gamma_{\rm strings}=\frac{g^2J_{\rm BH}}{\tau_{\rm BH}}\frac{\mathcal{F}_L}{4 \pi (m_{A'} d)^2}\theta(t_{{\rm MW}}-t_{\rm SH})
\eea
Where $J_{\rm BH}=a_* GM_{\rm BH}^2$ is the total angular momentum of the black hole and $t_{{\rm MW}}$ is the age of the Milky Way. If $t_{\rm SH}>t_{{\rm MW}}$, even a black hole produced at the genesis of the Milky Way will not have had enough time to reach the superheating threshold and thus will not produce any strings. To capture this, we have inserted the factor $\theta(t_{{\rm MW}}-t_{\rm SH})$.

Finally, we can consider the full ensemble of black holes in the galaxy. To do this we multiply Eq.~\ref{Eq: Multiple Bursts Rate} by $N_A$, the number of superradiantly active black holes and average over galactic distributions for black hole spin, mass, and distance. We use the same distributions used in Appendix C of~\cite{Arvanitaki:2014wva}. The number of active black holes is computed in appendix~\ref{Apdx: Bosenova Rate} as $N_A=\tau_{\rm BH}R_{\rm BHF}(1-e^{-t_{{\rm MW}}/\tau_{\rm BH}})$ where $R_{\rm BHF}$ is the rate of black hole formation in the galaxy. The final rate is given by 
\bea
\label{Eq: Final Bursts Rate}
\Gamma_{\rm strings}=\bigg< g^2J_{\rm BH}R_{\rm BHF}\frac{\mathcal{F}_L}{4 \pi (m_{A'} d)^2}(1-e^{-t_{{\rm MW}}/\tau_{\rm BH}})\theta(t_{{\rm MW}}-t_{\rm SH})\bigg>_{a_*,M_{\rm BH},d}
\eea

\subsection{Results}

\begin{figure}[t!]
\centering
\includegraphics[width=.8\linewidth]{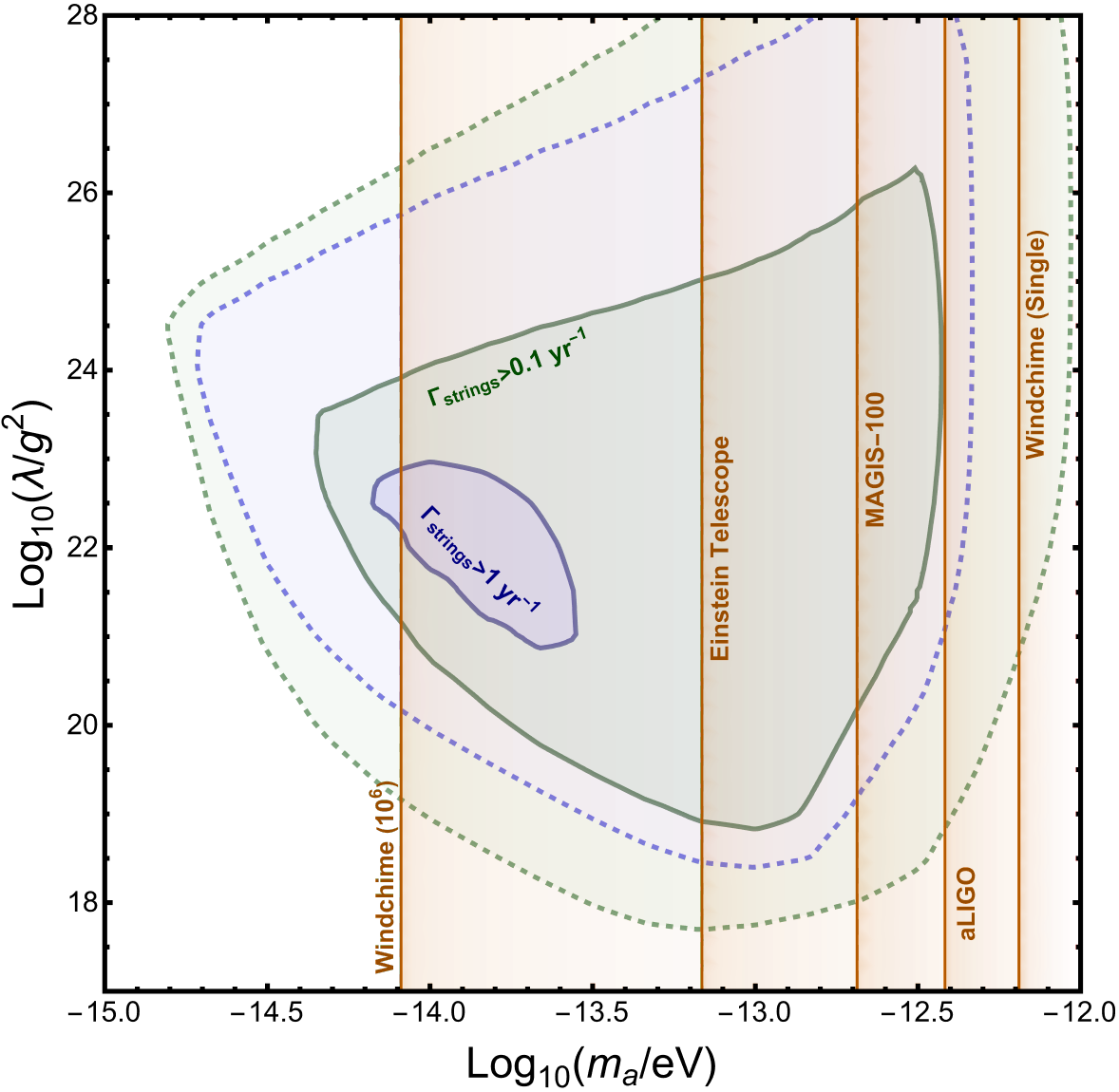}
\caption{A plot of $\Gamma_{\rm strings}$. The blue indicates the region where $\Gamma_{\rm strings}>1\; \rm yr^{-1}$ while red indicates $\Gamma_{\rm strings}>0.1\; \rm yr^{-1}$. The solid lines indicate the region where the average rates exceed these values while the dashed lines indicate the 1 sigma uncertainty in those regions. The golden lines and the shaded regions to the right indicate the masses, $m_{A'}$, at which the various experiments detect stings with ${\rm SNR}\geq1$. Windchime is taken to have detectors with sensitivity $S_n^{1/2}=10^{-13}$ m/s$^2$/$\sqrt{\text{Hz}}$, and $\Delta=0.05$ and regions are shown both for a single pair of detectors as well as for an array of $10^6$ detectors} 
\label{Fig: String Rates Contour}
\end{figure}

We numerically compute $\Gamma_{\rm strings}$ for various values of $m_{A'}$ and $\lambda/g^2$. We keep the coupling $g$ fixed at its maximal value from equivalence principle constraints to maximize the rate. We also compute the uncertainty in this rate both from a Poisson distribution of the random string arrivals and from the variance in the rate due to the spreads of the distributions of $M_{\rm BH}$, $d$ and $a_*$. The dominant contribution to this uncertainty is from the variance in the black hole distance since closer black holes will have significantly higher rates due to the $\Gamma_{\rm strings}\propto d^{-2}$ scaling. Figure~\ref{Fig: String Rates Contour} shows a contour plot of $\Gamma_{\rm strings}$. The regions enclosed by the solid blue and green lines represent the regions of parameter space where we expect $\Gamma_{\rm strings}>1 \;\rm yr^{-1}$ and $\Gamma_{\rm strings}>0.1 \;\rm yr^{-1}$ on average, respectively. The dashed lines indicate the one-sigma uncertainty in these regions. To indicate the regions of this parameter space where we may be able to detect a passing string, we include golden lines at the mass $m_{A'}$ where the ${\rm SNR}$ from a given experiment exceeds 1 with ${\rm SNR}>1$ to the right of all these lines, as indicated by the shading.

We can see from figure~\ref{Fig: String Rates Contour}, that there is potentially a large range of parameter space where strings can both be produced abundantly enough to reach Earth and with masses that will be detectable to present and future experiments. aLIGO is the only currently operating experiment of the bunch and has a small overlap with the confidence regions of both the $\Gamma_{\rm String}=1$ yr$^{-1}$ and $\Gamma_{\rm String}=0.1$ yr$^{-1}$ contours meaning it could have a chance to observe a string during its runtime. These chances of observation improve as we consider future experiments with MAGIS$-100$ and Einstein Telescope further overlapping with these regions. Finally, in the far future, a Windchime configuration with $10^6$ accelerometers, would be able to observe nearly all of the possible parameter space where strings could reach earth.

\section{Conclusion}

In this paper, we discussed in more detail the subsequent evolution of dark photon string network after the superheated phase transition discovered in~\cite{East:2022ppo,East:2022rsi}, as well as the corresponding novel observable signatures.  Spinning black holes periodically emit bursts of strings loops, whose subsequent emission of gravitational waves and collision with detectors on Earth are observable at both current and future detectors.

On the gravitational wave side, the smoking gun signatures of the string bosenova events are {\it repeating} (and most likely periodic) gravitational wave bursts with similar spectral properties among bursts. These bursts can have a wide range of peak frequency, depending mainly on the unknown mass of the dark photon and the parameter $\lambda/g^2$. However, for any particular dark photon mass, the peak frequency shall be roughly the same for all black holes that are experiencing periodic string bosenova. The repeating string bosenova can occur around black holes of all masses from stellar mass black holes to supermassive black holes, and the bosenova signal can show up in gravitational wave detectors that are operating at all frequencies from $ \sim {\rm nHz}$ to 10 ${\rm MHz}$. The observational prospect at various high frequency gravitational wave detectors are of particular interest, as this is one of the only sources for strong gravitational wave with frequencies that are much larger than $\sim {\rm kHz}$, that are not already tightly constrained by the $N_{\rm eff}$ bound from CMB observations.

On the direct detection side, the bosenova events produce topological defects that can interact with ground based detectors~\cite{Pospelov:2012mt}. The novelty is that this interaction is coupling strength independent. The interaction between dark photon strings with a radius of $1/m_{A'} \gtrsim 10\,{\rm km}$ and lengths longer than the Earth radius leads to predictable strain (acceleration) as a function of frequency, depending only on the dark photon mass. Using light $B-L$ gauge boson as an example, we highlight the EP violating nature of this acceleration, and compare the prospects for detection at gravitational wave detectors and EP violation tests. With current and future detectors, we can expect detectable signal with rates larger than $1/{\rm yr}$ for dark photon with masses between $10^{-14}\,{\rm eV}$ and $10^{-12}\,{\rm eV}$ in current and future detectors.

The results we presented in the paper relied on assumptions about the evolution of the string network, parameterized by two distributions functions: $F_s(l,t)$ which parameterize the length distribution of the dark photon string as a function of time, and $F_r(k,t)$ which parameterize the distribution of radiation energy density as a function of energy (momentum) and time. Input from simulations of a string network with the superheated phase transition as the initial condition will be critical for determining the properties of these distribution functions as the string network expands~\footnote{The presence of the central black hole will quickly lose importance as the coupled system of dark photon strings and radiation expand to sizes much larger than the superradiance cloud.}.  The observational properties of the gravitational wave signatures are strongly dependent on these distributions and properly establishing their behavior will be important input at gravitational wave detectors.

In this paper, we discussed the observational consequences of string bosenova in gravitational wave detectors and accelerometers. These observational consequences only depended on the dark photon mass, and the parameters in the Higgs mechanism that generates this mass. Both classes of signatures do not depend directly on the strength of coupling between the dark photon and the Standard Model particles, a consequence of the topological nature of the object we are searching for.  Additional signatures of these dark photon strings can be induced by considering additional interactions between the dark photon and the Standard Model particles, for example, kinetic mixing. We leave these signatures to future studies.

\section*{Acknowledgments}
We thank Mustafa Amin, Asher Berlin, Will East, Andrew Geraci, Ed Hardy, Maxim Pospelov and Nils Siemonsen for helpful discussion.  
DB, AH and CR are supported by NSF grant PHY-2210361 and the Maryland Center for Fundamental Physics.
JH is supported by the National Science and Engineering Research Council through a Discovery grant.
Research at Perimeter Institute is supported in part by the Government of Canada through the Department of Innovation, Science and Economic Development Canada and by the Province of Ontario through the Ministry of Research, Innovation and Science.
\appendix

\section{Rate of Bosenova Events}\label{Apdx: Bosenova Rate}

Here, we estimate the Bosenova rate $\Gamma_{{\rm MW}}$ in the Milky Way. The rate of these events is the product of the number of "active" black holes, $N_A$, and the bosenova rate $t_{\rm SH}^{-1}$.
\bea \label{Eq: MW Bosenova rate}
\Gamma_{{\rm MW}} = \frac{N_A}{t_{\rm SH}}
\eea
Black holes are formed with a rate $R_{\rm BHF}$ in the Milky Way and exist in their superradniently active phase for $\tau_{\rm BH}$ which allows us to write an equation for the number of active black holes 
\bea
\label{Eq: Diff Eq for NA}
\dot{N}_A = R_{\rm BHF} - \frac{N_A}{\tau_{\rm BH}}
\eea
which can easily be solved to find 
\bea
\label{Eq: NA}
N_A =\tau_{\rm BH}R_{\rm BHF}(1-e^{-t_{{\rm MW}}/\tau_{\rm BH}})
\eea
where $t_{{\rm MW}}$ is the age of the Milky Way. From this one can easily compute $\Gamma_{{\rm MW}}$ in the two limits $t_{{\rm MW}}\gg \tau_{\rm BH}$ and $t_{{\rm MW}}\ll \tau_{\rm BH}$
\bea
\label{Eq:limitsofGammaMW}
\Gamma_{{\rm MW}}=\frac{R_{\rm BHF}}{\mathcal{F}_s} \quad \text{if} \quad t_{{\rm MW}}\gg \tau_{\rm BH} \\
\Gamma_{{\rm MW}}=\frac{t_{{\rm MW}}}{t_{\rm SH}}R_{\rm BHF}\quad \text{if} \quad t_{{\rm MW}}\ll \tau_{\rm BH}
\eea

\section{Analytic expressions for characteristic strains} \label{Appendix: exact results}

In this appendix, we provide the analytic results for characteristic strains shown in sections \ref{sec: GW tightly coupled}, and \ref{sec: GW decoupled}, subject to all of the approximations described in those sections.

\subsection{Tightly coupled regime}

Here, we present results for the characteristic strain of the signal produced during the tightly coupled regime up to contributions scaling as powers of $\l \frac{\Lt}{L_{\rm tot}}\r$.  The results for various values of $q$ are as follows:\\
For $q>-1$\\ 
\bea
h_c(f) &=& \frac{ \sqrt{2 \Gamma_{\rm GW} L_{\rm tot} \Lt}   G \mu }{ \pi d } \sqrt{\frac{2+q}{1+q}} \nonumber \\
&\cdot & \begin{cases}  
\sqrt{1 - \l   \frac{f L_{\rm min}(0)}{2 }  \r^{\frac{2+2q}{3}}}  \l \frac{L_{\rm max}(0)}{\Lt} \r^{\frac{1}{2}} \l \frac{2}{f L_{\rm max}(0)} \r^{\frac{2+q}{2}}  & \frac{2}{L_{\rm min} (0)} > f > 2 m_{A'} \\
\sqrt{ 1 - \l \frac{L_{\rm max}(0)}{\Lt}\r^{1+q }}  \l \frac{L_{\rm max}(0)}{\Lt} \r^{\frac{1}{2}} \l \frac{2}{f L_{\rm max}(0)} \r^{\frac{2+q}{2}}  & 2 m_{A'} > f > \frac{2}{L_{\rm max} (0)}
\\
\sqrt{ 1 - \l \frac{2}{f \Lt} \r^{1+q}}    \l \frac{2}{f \Lt} \r^{\frac{1}{2}}  & \frac{2}{L_{\rm max} (0)} > f > \frac{2}{\Lt} 
\\
0 & f > \frac{2}{L_{\rm min} (0)} \text{ or } f < \frac{2}{\Lt}, 
\end{cases}  
\eea
\noindent for $q=-1$
\bea
h_c(f) = \frac{ \sqrt{2 \Gamma_{\rm GW} L_{\rm tot} \Lt}   G \mu }{ \pi d } \l \frac{2}{f \Lt} \r^{\frac{1}{2}}  \cdot \begin{cases}  
\sqrt{\frac{2}{3}\ln \l   \frac{2 }{f L_{\rm min}(0)}  \r}     & \frac{2}{L_{\rm min} (0)} > f > 2 m_{A'} \\
\sqrt{ \ln \l \frac{\Lt}{L_{\rm max}(0)}\r}   & 2 m_{A'} > f > \frac{2}{L_{\rm max} (0)}
\\
\sqrt{ \ln \l \frac{f \Lt}{2} \r}      & \frac{2}{L_{\rm max} (0)} > f > \frac{2}{\Lt} 
\\
0 & f > \frac{2}{L_{\rm min} (0)} \text{ or } f < \frac{2}{\Lt} ,
\end{cases}
\eea

\noindent for $q=-2$ 
\bea
h_c(f) = \frac{ 2\sqrt{ \Gamma_{\rm GW} } L_{\rm tot}  G \mu }{ \pi d }  \cdot \begin{cases}
\sqrt{- {\rm li} \l \l \frac{2}{f L_{\rm tot}} \r^{\frac{2}{3}} \r + {\rm li} \l \frac{L_{\rm max}(0)}{ L_{\rm tot}}  \r }      & \frac{2}{L_{\rm min} (0)} > f > 2 m_{A'} 
\\
\sqrt{- {\rm li}  \l \frac{\Lt}{ L_{\rm tot}}  \r + {\rm li} \l  \frac{L_{\rm max}(0)}{ L_{\rm tot}}  \r }    & 2 m_{A'} > f > \frac{2}{L_{\rm max} (0)}
\\
\sqrt{- {\rm li}  \l \frac{\Lt}{ L_{\rm tot}}  \r + {\rm li} \l  \frac{2}{ f L_{\rm tot}}  \r }    & \frac{2}{L_{\rm max} (0)} > f > \frac{2}{\Lt} 
\\
0 & f > \frac{2}{L_{\rm min} (0)} \text{ or } f <  \frac{2}{\Lt} ,
\end{cases}   
\eea
where $ {\rm li}(x) = \int\limits^{x}_0 \frac{d t}{\ln t}$. Finally,
\noindent for $q<-2$

\bea
h_c(f) &=& \frac{ \sqrt{2 \Gamma_{\rm GW} L_{\rm tot} \Lt}   G \mu }{ \pi d } \sqrt{\frac{4+2q}{4+3 q}} \nonumber\\
&\cdot& \begin{cases}
\sqrt{1 - \l \frac{f L_{\rm min} (0)}{2 } \r^{-\frac{4}{3}-q} }  \l \frac{f}{2 \mA} \r^{-\frac{1}{3}}   & \frac{2}{L_{\rm min} (0)} > f > 2 m_{A'} 
\\
\sqrt{  1 -  \l m_{A'} L_{\rm min} (0)  \r^{-\frac{4}{3}-q} }   \l \frac{f}{2 m_{A'}} \r^{\frac{-2-q}{2}} & 2 m_{A'} > f > \frac{2}{L_{\rm max} (0)}
\\
\sqrt{ 1 -  \l \frac{2}{f \Lt } \r^{-2-\frac{3}{2}q}}  \l \frac{f}{2 m_{A'}} \r^{\frac{-2-q}{2}}  & \frac{2}{L_{\rm max} (0)} > f > \frac{2}{\Lt} 
\\
0 & f > \frac{2}{L_{\rm min} (0)} \text{ or } f <  \frac{2}{\Lt} .
\end{cases}   
\eea

As mentioned in the main text, much of the scaling behavior of these solutions can be understood with back of the envelope estimates.  From Eq.~\ref{eq: rho and strain}, we have $h_c^2 \propto f^{-1} d \rho_{\rm GW}/d\log f \propto f^{-1} n_{\rm eff}(f) t_{\rm eff} (f) $, where $n_{\rm eff}(f)$ is effectively the number of strings emitting with frequency $f$ and $t_{\rm eff} (f)$ is how long they are emitting for at that length.  
There are two populations of strings that emit gravitational waves, the typical strings and the tails of the distribution.  The typical strings have size $L_{\rm max}$ ($L_{\rm min}$) if the strings are IR (UV) dominated, which appear on Figures~\ref{fig: tightly coupled spectrum example}, \ref{fig: decoupled spectrum example},~\ref{fig: combined spectrum example} and~\ref{fig: spectrum contributions example} as the peak of the spectrum, while the spectral index $q$ determines the corresponding powerlaw falloff in the UV(IR).  In what follows we give a heuristic argument for the scaling behavior of Eq. \eqref{eq: hc SI IR} and Eq. \eqref{eq: hc SI UV}.

For the part of the spectrum driven by the typical strings, the number of contributing strings is $n_{\rm eff}(f) \sim L_{\rm tot} / l \sim f$, since strings only emit at a frequency $f$ when their length $l$ is $\sim 1/f$. Thus the slope in this regime is solely determined by how long strings contribute to the given frequency $h_c \propto \sqrt{t_{\rm eff} (f)}$.  Since IR (UV) dominated distributions have their time dependent string lengths $l \sim t_{\rm eff}$ ($l \sim t_{\rm eff}^{3/2}$),  we get $h_c \propto f^{-1/2}$ for distributions dominated by long strings and $h_c \propto f^{-1/3}$ for distributions dominated by short strings. In either case, the peak frequency is located when $t_{\rm eff}(f) = t_{\rm trans}$, which is why the values of peak strain $h_{c,{\rm peak}}$ are approximately distribution independent, as evident in Eq. \eqref{eq: hc SI IR}, Eq. \eqref{eq: hc SI UV}, Fig.~\ref{fig: tightly coupled spectrum example}, and Fig.~\ref{fig: spectrum contributions example} (colored dotted lines).

The part of the spectrum associated with the tails of the distribution are being dominantly emitted all at the same time ($t \sim L_{\rm max}(0)$ for IR dominated string distributions and $t \sim \Lt$ for UV dominated string distributions) and hence the slope is $h_c \propto f^{-1/2} \sqrt{n_{\rm eff}(f)}$. Since $n_{\rm eff}(f) \sim l F_s(l, t) \propto l^{1+q} \propto f^{-1-q}$, we get $h_c \propto f^{-(2+q)/2}$ for both short and long string dominated distributions.

\subsection{Decoupled regime}

Now, we present results for the characteristic strain of the signal produced when strings are decoupled. The results for various values of $q$ are the following:\\
For $q>-1$\\

\bea
h_{c, {\rm ssc}}(f) &=& \frac{ \sqrt{2 \Gamma_{\rm GW} }L_{\rm tot}   G \mu }{ \pi d \sqrt{ 1 - \l  \frac{l_{\rm min} (L_{\rm tot})}{ L_{\rm tot}} \r^{q+2}}} \sqrt{\frac{2+q}{1+q}} \nonumber \\
&\cdot& \begin{cases}  
\sqrt{1 - \l   \frac{f l_{\rm min}(0)}{2 }  \r^{1+q}}  \l \frac{\Lt}{L_{\rm tot}} \r^{\frac{1}{2}} \l \frac{2}{f \Lt} \r^{\frac{2+q}{2}}  & \frac{2}{l_{\rm min} (0)} > f > \frac{2}{l_{\rm min} (L_{\rm tot})} \\
\sqrt{ 1 - \l \frac{\Lt}{L_{\rm tot}}\r^{1+q }}  \l \frac{\Lt}{L_{\rm tot}} \r^{\frac{1}{2}} \l \frac{2}{f \Lt} \r^{\frac{2+q}{2}}  & \frac{2}{l_{\rm min} (L_{\rm tot})} > f > \frac{2}{\Lt}
\\
\sqrt{ 1 - \l \frac{2}{f L_{\rm tot}} \r^{1+q}}    \l \frac{2}{f L_{\rm tot}} \r^{\frac{1}{2}}  & \frac{2}{\Lt} > f > \frac{2}{L_{\rm tot}} 
\\
0 & f > \frac{2}{l_{\rm min} (0)} \text{ or } f < \frac{2}{L_{\rm tot}}, 
\end{cases}  
\eea

\bea
h_{c, {\rm dec}}(f) &=& \frac{ \sqrt{2 L_{\rm tot}^3 \mA}  G \mu }{ \pi d \sqrt{ 1 - \l  \frac{l_{\rm min} (L_{\rm tot})}{ L_{\rm tot}} \r^{q+2}} } \sqrt{\frac{\Gamma_{\rm GW}}{\Gamma_{\rm DP}}} \sqrt{\frac{2+q}{1+q}} \nonumber\\
&\cdot& \begin{cases}  
\sqrt{   1 - \l  \frac{2}{f L_{\rm tot}} \r^{q+1}  }  \l \frac{2}{f L_{\rm tot}} \r^{\frac{3}{2}}  & \frac{2}{l_{\rm min} (L_{\rm tot})} > f > \frac{2}{L_{\rm tot}} 
\\
\sqrt{  1 - \l  \frac{l_{\rm min} (L_{\rm tot})}{ L_{\rm tot}} \r^{q+1}  }  \l \frac{2}{f L_{\rm tot}} \r^{\frac{3}{2}}  & 2 m_{A'} > f > \frac{2}{l_{\rm min} (L_{\rm tot})} 
\\
\sqrt{\frac{ 1 - \l  \frac{l_{\rm min} (L_{\rm tot})}{ L_{\rm tot}} \r^{q+1}}{m_{A'} L_{\rm tot}} }   \l \frac{2}{f L_{\rm tot}} \r  & f > 2 m_{A'}
\\
0 & f < \frac{2}{L_{\rm tot}},
\end{cases} 
\eea
for $q=-1$
\begin{eqnarray}    
h_{c, {\rm ssc}}(f) &=& \frac{ \sqrt{2 \Gamma_{\rm GW}  } L_{\rm tot}   G \mu }{ \pi d \sqrt{ 1 -  \frac{l_{\rm min} (L_{\rm tot})}{ L_{\rm tot}} }} \l \frac{2}{f L_{\rm tot}} \r^{\frac{1}{2}}  \cdot \begin{cases}  
\sqrt{\ln \l   \frac{2 }{f l_{\rm min}(0)}  \r}     & \frac{2}{l_{\rm min} (0)} > f > \frac{2}{l_{\rm min} (L_{\rm tot})} \\
\sqrt{ \ln \l \frac{L_{\rm tot}}{\Lt}\r}   & \frac{2}{l_{\rm min} (L_{\rm tot})} > f > \frac{2}{\Lt}
\\
\sqrt{ \ln \l \frac{f L_{\rm tot}}{2} \r}      & \frac{2}{\Lt} > f > \frac{2}{L_{\rm tot}} 
\\
0 & f > \frac{2}{l_{\rm min} (0)} \text{ or } f < \frac{2}{L_{\rm tot}},
\end{cases} \\
h_{c, {\rm dec}}(f) &=& \frac{ \sqrt{2 L_{\rm tot}^3 \mA} G \mu }{ \pi d \sqrt{ 1 -  \frac{l_{\rm min} (L_{\rm tot})}{ L_{\rm tot}} } } \sqrt{\frac{\Gamma_{\rm GW}}{\Gamma_{\rm DP}}}  \nonumber\\
&\cdot& \begin{cases}  
\sqrt{  \ln \l  \frac{2}{f L_{\rm tot}} \r  }  \l \frac{2}{f L_{\rm tot}} \r^{\frac{3}{2}}  & \frac{2}{l_{\rm min} (L_{\rm tot})} > f > \frac{2}{L_{\rm tot}} 
\\
\sqrt{  \ln \l  \frac{l_{\rm min} (L_{\rm tot})}{ L_{\rm tot}} \r  }  \l \frac{2}{f L_{\rm tot}} \r^{\frac{3}{2}}  & 2 m_{A'} > f > \frac{2}{l_{\rm min} (L_{\rm tot})} 
\\
\sqrt{ \ln \l  \frac{l_{\rm min} (L_{\rm tot})}{ L_{\rm tot}} \r} \l m_{A'} L_{\rm tot} \r^{-\frac{1}{2}}  \l \frac{2}{f L_{\rm tot}} \r  & f > 2 m_{A'}
\\
0 & f < \frac{2}{L_{\rm tot}},
\end{cases} 
\end{eqnarray}
for $q=-2$
\begin{eqnarray}
h_{c, {\rm ssc}}(f) &=& \frac{ 2\sqrt{ \Gamma_{\rm GW}  }L_{\rm tot}   G \mu }{ \pi d \sqrt{ \ln \l  \frac{ L_{\rm tot}}{l_{\rm min} (L_{\rm tot})} \r} }  \cdot \begin{cases}
\sqrt{1 - \frac{f l_{\rm min} (0)}{2}} \l  \frac{2}{f l_{\rm min} (L_{\rm tot})} \r^{\frac{1}{2}}   & \frac{2}{l_{\rm min} (0)} > f > \frac{2}{l_{\rm min} (L_{\rm tot})} 
\\
\sqrt{1 -  \frac{\Lt}{ L_{\rm tot}}    }    & \frac{2}{l_{\rm min} (L_{\rm tot})} > f > \frac{2}{\Lt}
\\
\sqrt{1 -  \frac{2}{f L_{\rm tot}}  }    & \frac{2}{\Lt} > f > \frac{2}{L_{\rm tot}}  
\\
0 & f > \frac{2}{l_{\rm min} (0)} \text{ or } f < \frac{2}{L_{\rm tot}} ,
\end{cases}  \\ 
h_{c, {\rm dec}}(f) &=& \frac{ \sqrt{2 L_{\rm tot}^3 \mA} G \mu }{ \pi d \sqrt{ \ln \l  \frac{ L_{\rm tot}}{l_{\rm min} (L_{\rm tot})} \r} } \sqrt{\frac{\Gamma_{\rm GW}}{\Gamma_{\rm DP}}}  \nonumber\\
&\cdot& \begin{cases}  
\sqrt{  1 -   \frac{2}{f L_{\rm tot}}   }  \l \frac{2}{f L_{\rm tot} } \r  & \frac{2}{l_{\rm min} (L_{\rm tot})} > f > \frac{2}{L_{\rm tot}}
\\
\sqrt{  1 -   \frac{l_{\rm min} (L_{\rm tot})}{ L_{\rm tot}}   } \l \frac{ L_{\rm tot} }{l_{\rm min} (L_{\rm tot})} \r^{\frac{1}{2}} \l \frac{2}{f L_{\rm tot} } \r^{\frac{3}{2}}  & 2 m_{A'} > f > \frac{2}{l_{\rm min} (L_{\rm tot})}
\\
\sqrt{ \frac{1 - \frac{l_{\rm min} (L_{\rm tot})}{ L_{\rm tot}} }{\mA l_{\rm min} (L_{\rm tot})}} \l \frac{2}{f L_{\rm tot} } \r  & f > 2 m_{A'}
\\
0 & f < \frac{2}{L_{\rm tot}},
\end{cases}  
\end{eqnarray}
for $q<-2$
\bea
h_c(f) &=& \frac{ \sqrt{2 L_{\rm tot} m_{A'}^{-1}} G \mu }{ \pi d \sqrt{ 1 - \l  \mA \Lt \r^{q+2}}} \sqrt{\frac{\Gamma_{\rm GW}}{\Gamma_{\rm DP}}} \sqrt{\frac{2+q}{1+q}} \nonumber\\
&\cdot& \begin{cases}  
\sqrt{  1 - \l  \frac{f \Lt}{2} \r^{q+1}  }  \l \frac{f }{2 m_{A'}} \r^{\frac{1}{2}}  & 2 m_{A'} > f > \frac{2}{\Lt} 
\\
\sqrt{ 1 - \l  m_{A'} \Lt \r^{q+1}}  \l \frac{f }{2 m_{A'}} \r^{-1}  & f > 2 m_{A'}
\\
0 & f < \frac{2}{\Lt}.
\end{cases}  
\eea

Different scalings of the spectra can be understood similarly to the tightly coupled case. Again, the characteristic strain depends only on $t_{\rm eff} (f)$ and $n_{\rm eff}(f)$, through $h_c^2 \propto f^{-1} d \rho_{\rm GW}/d\log f \propto f^{-1} n_{\rm eff}(f) t_{\rm eff} (f) $. 

\begin{figure*}[t] 
\centering
\begin{minipage}{.48\textwidth}
    \centering
    \includegraphics[width=0.995\linewidth]{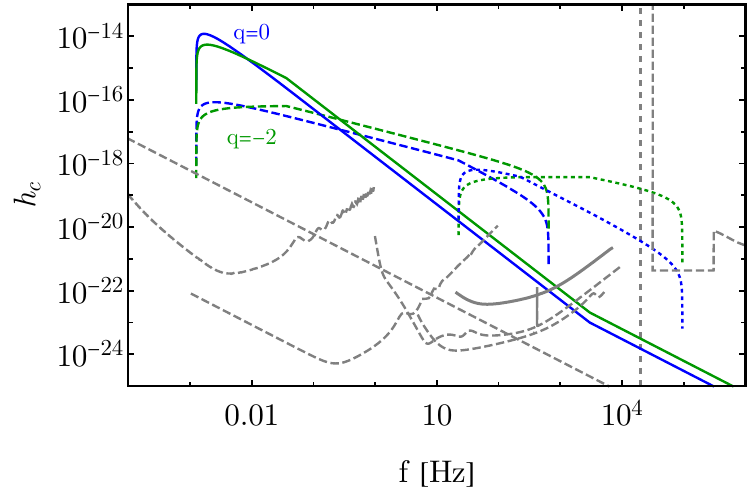}
\end{minipage}%
\hfill
\begin{minipage}{.48\textwidth}
    \centering
    \includegraphics[width=0.995\linewidth]{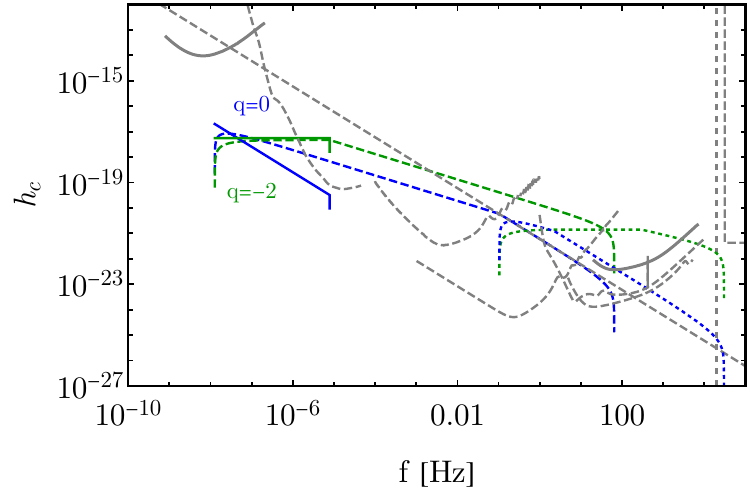}
\end{minipage}
\vskip\baselineskip

\begin{minipage}{.48\textwidth}
    \centering
    \includegraphics[width=0.995\linewidth]{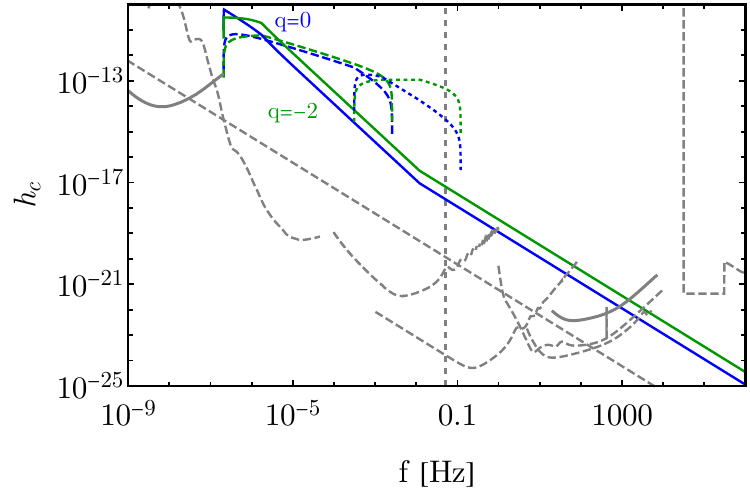}

\end{minipage}%
\hfill
\begin{minipage}{.48\textwidth}
    \centering
    \includegraphics[width=0.995\linewidth]{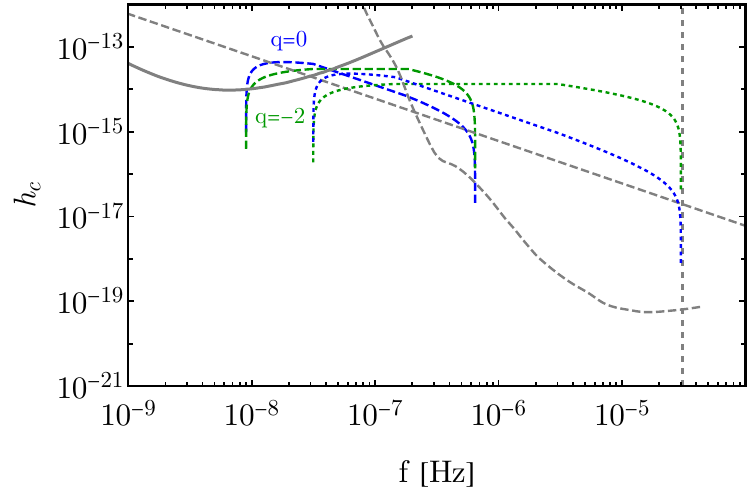}
\end{minipage}
\caption{Comparison of strain spectra generated during tightly coupled (colored dotted lines), efficient string-string interaction (colored dashed lines), and decoupled (colored solid lines) regimes for parameter choices the same as in Figure~\ref{fig: tightly coupled spectrum example} and different spectral index $q$. If the duration of the signal is longer than $20$ years, we include only the first $20$ years of the signal.    
}
    \label{fig: spectrum contributions example}
\end{figure*}

Let us first discuss IR dominated string distributions.  The GW emission coming from when string-string interactions are efficient (colored dashed lines in Fig.~\ref{fig: spectrum contributions example}) proceeds exactly like the tightly coupled regime and hence its behavior is the same up to various substitutions.  This explains why Eq. \eqref{eq: hc SI IR} and Eq. \eqref{eq: hc DPdecoupled IR} are nearly identical.
The other part of the IR dominated string distribution is Eq. \eqref{eq: hc decoupled IR}, the GW emission from the long string (colored solid lines in Fig.~\ref{fig: spectrum contributions example}). Since there is only the one string, we have $h_c \sim \sqrt{\tau(f)/f}$, giving the scalings found in Eq. \eqref{eq: hc decoupled IR}.

For UV dominated string distributions, a similar situation prevails in Eq. \eqref{eq: hc decoupled uv}.  For frequencies larger than $f > \mA$, GW emission is dominated by the typical string whose starting length is $l \sim 1/\mA$ giving $h_c \sim \sqrt{\tau(f)/f} \sim 1/f$.  
For frequencies smaller than $\mA$, GW emission is dominated by when $l_{\rm min}(t) \sim 1/f$, namely by when the strings emitting that particular frequency are just decoupling from the rest of the string distribution.  Combining $\tau \sim \mA l_{\rm min}^2$ and $n_{\rm eff}(f) \sim L_{\rm tot}(t)/l_{\rm min} \sim l_{\rm min}^{-4}$, we find the low frequency scaling of Eq. \eqref{eq: hc decoupled uv}.

\section{Dependence of String Arrival Rate on String Length Distribution}
\label{Apdx: FL computation}
In this appendix we examine the effect that the string length distribution at the transition to the free streaming era, $F_s(l_0,t_{\rm trans})$ has on the string arrival rate on earth. To quantify this we will look at the $\mathcal{F}_L(t)$ the fraction of total length remaining after time $t$ in the free streaming regime defined just below Eq.~\ref{Eq: Single Burst Rate}. If one makes the somewhat crude simplification $l(l_0,t)= l_0\theta(\tau-t)$ where $\tau$ is the lifetime of the string given in Eq.~\ref{Eq: String Lifetime}, one can compute an analytic expression $\mathcal{F}_L(d)$ for some power law string distribution $F_s(l_0)\propto l_0^q$ for $q\neq 2$
\bea
\label{Eq: Analytic Energy Fraction}
\mathcal{F}_L=\frac{1-\l \alpha \l \lambda/g^2\r^{-1/3}\sqrt{2\Gamma_D m_{A'} d}\r^{q+2}}{1-\l \alpha \l \lambda/g^2\r^{-1/3}\r^{q+2}}
\eea
Using the facts that $\alpha \l \lambda/g^2\r^{-1/3}\sqrt{2\Gamma_D m_{A'} d}<1$ and $\alpha \l \lambda/g^2\r^{-1/3}\ll 1$ we can take both the $q>2$ and $q<2$ limits. 
\bea
\mathcal{F}_L\approx 1-\l \alpha \l \lambda/g^2\r^{-1/3}\sqrt{2\Gamma_D m_{A'} d}\r^{q+2} \quad \text{if}\quad q>-2\\
\mathcal{F}_L\approx\l 2\Gamma_D m_{A'} d\r^{q/2+1}\ll 1\quad \text{if}\quad q<-2
\eea 
 For $q>-2$, $\mathcal{F}_L$ decreases like $d^{q/2+1}$ from 1 to 0. Conversely for $q<-2$ since $m_{A'} d\gg1 $, we find that $\mathcal{F}_L\ll 1$. Physically, this is because, for $q<-2$, most of the length is stored in strings of minimal length and thus cannot survive the journey while for $q>-2$, most of the length is stored in strings with initial length on the order of the size of the network and thus can survive the journey provided $d$ is not too long.

\begin{figure}[h!]
\centering
\includegraphics[width=.7\linewidth]{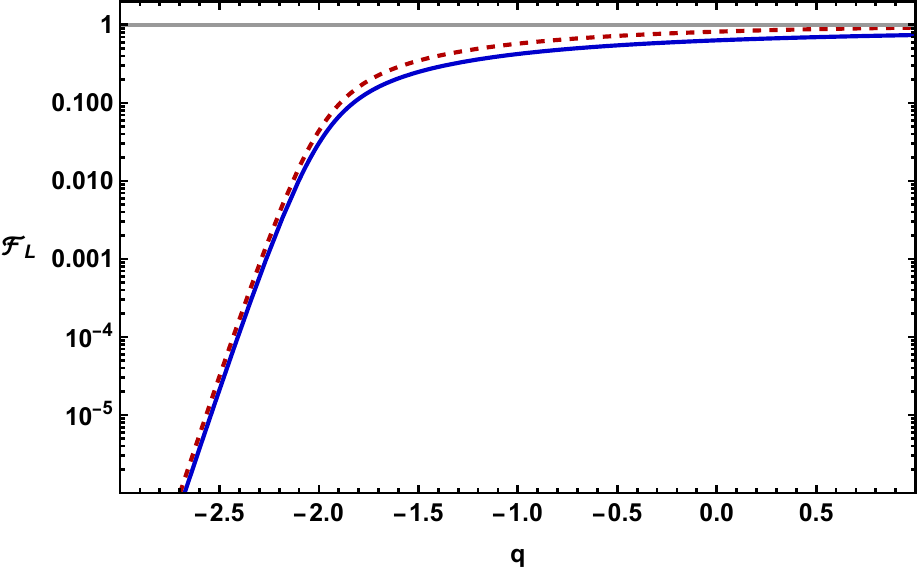}
\caption{A plot of the length fraction $\mathcal{F}_L$ as a function of $q$ for $d=8$ kpc, $m_{A'}=10^{-13}$ eV, and $\lambda/g^2=10^{19}$. The solid blue line represents the exact numerical computation of $\mathcal{F}_L$ while the dashed red line indicates the analytic approximation given in Eq.~\ref{Eq: Analytic Energy Fraction}. Although the exact numerical value of the two don't align exactly, they both exhibit the same  roughly constant behavior for $q>-2$ and a steep drop off for $q<-2$.} 
\label{Fig: FL Plot}
\end{figure}

 In practice, one can compute $l(l_0,t)$ exactly from Eq.~\ref{Eq: Total Power} in the $G\mu\ll1$ limit and can then numerically compute $\mathcal{F}_L$. Figure~\ref{Fig: FL Plot} shows a plot of both the exact numerical computation analytic approximation of $\mathcal{F}_L(q)$ for $d=8$ kpc, $m_{A'}=10^{-13}$ eV, and $\lambda/g^2=10^{19}$. In both we find that the length fraction is roughly constant and $\OO(1)$ for $q>-2$ with a steep drop off at $q<-2$ as expected. 

\bibliography{biblio}{}

\bibliographystyle{JHEP}

\end{document}